\documentclass[%
notitlepage,
superscriptaddress,
nofootinbib,
amsmath,
amssymb,
aps,
twocolumn
]{revtex4-1}
\usepackage{graphicx}% Include figure files
\usepackage{dcolumn}% Align table columns on decimal point
\usepackage{bm}% bold math
\usepackage{mathtools}
\usepackage{amsmath}
\usepackage{amssymb}
\usepackage[hidelinks]{hyperref}% add hypertext capabilities
\usepackage{setspace}% fixes weird extra space that got inserted into the appendix
\usepackage[dvipsnames, usenames]{xcolor}

\begin{document}

\title{Conservative finite volume scheme for first-order viscous relativistic
hydrodynamics}

\author{Alex Pandya}
\email{apandya@princeton.edu}
\affiliation{Department of Physics, Princeton University, Princeton, New Jersey
08544, USA.}
\affiliation{Princeton Gravity Initiative, Princeton University, Princeton, NJ 08544, USA}
\author{Elias R. Most}
\email{emost@princeton.edu}
\affiliation{Princeton Gravity Initiative, Princeton University, Princeton, NJ 08544, USA}
\affiliation{Princeton Center for Theoretical Science, Princeton University, Princeton, NJ 08544, USA}
\affiliation{School of Natural Sciences, Institute for Advanced Study, Princeton, NJ 08540, USA}
\author{Frans Pretorius}
\email{fpretori@princeton.edu}
\affiliation{Department of Physics, Princeton University, Princeton, New Jersey
08544, USA.}
\affiliation{Princeton Gravity Initiative, Princeton University, Princeton, NJ 08544, USA}

\date{\today}

\begin{abstract}
We present the first conservative finite volume numerical scheme for the
causal, stable relativistic Navier-Stokes equations developed by Bemfica,
Disconzi, Noronha, and Kovtun (BDNK).  BDNK theory has arisen very recently as
a promising means of incorporating entropy-generating effects (viscosity,
heat conduction) into relativistic fluid models, appearing as a possible
alternative to the so-called M\"uller-Israel-Stewart (MIS) theory successfully
used to model quark-gluon plasma.  
The major difference between the two lies in the structure of the system of PDEs: BDNK theory only has a set of conservation laws, whereas MIS also includes a set of evolution equations for its dissipative degrees of freedom.  The simpler structure of the BDNK PDEs in this respect allows for
rigorous proofs of stability, causality, and hyperbolicity in full generality 
which have as yet been impossible for MIS.  To capitalize on these
advantages, we present the first fully conservative multi-dimensional fluid solver for the BDNK equations
suitable for physical applications. The scheme includes a flux-conservative
discretization, non-oscillatory reconstruction, and a central-upwind numerical flux,
and is designed to smoothly transition to a high-resolution shock-capturing perfect
fluid solver in the inviscid limit.
We assess the robustness of our new method in a series of
flat-spacetime tests for a conformal fluid, and provide a detailed
comparison with previous approaches of Pandya \& Pretorius (2021) \cite{Pandya_2021}.
\end{abstract}

\maketitle

\section{Introduction} \label{sec:introduction}
Relativistic hydrodynamics is a general framework based on the notion that many
substances, even if governed by vastly different physics on small
spatiotemporal scales, may be well understood on sufficiently large scales by
appealing to thermodynamics and conservation laws for the local energy,
momentum, and baryon number \cite{Kovtun_2012}. This framework has resulted in
the development of successful fluid models of even exotic substances, such as
the quark-gluon plasma (QGP) produced in collisions of heavy ions, black hole
accretion flows, 
and the matter composing neutron stars. 
Experimental breakthroughs in studying the QGP, in particular, have spurred
significant growth in the theoretical understanding of relativistic fluids, as
the Relativistic Heavy Ion Collider (RHIC) is now sufficiently sensitive to
observe phenomena beyond the scope of ideal (non-dissipative) hydrodynamics
\cite{Romatschke_2019}.  Likewise, there are indications that similar phenomena
may become relevant in modeling astrophysical sources for the next generation
of telescopes and gravitational-wave observatories \cite{Alford_2018}
\cite{Most2021bulk}, further motivating theoretical focus on extending
relativistic hydrodynamics beyond thermodynamic equilibrium.

The modern interpretation of hydrodynamics views it as the expansion of a given microscopic physical theory (such as, for example, kinetic theory) for small values of the Knudsen number, $\textnormal{Kn} \equiv \ell/L \ll 1$, which gives the ratio of the microscopic ($\ell$) and macroscopic ($L$) lengthscales characterizing the system of interest.  The primary advantage a fluid model provides lies in the notion that many of the microscopic degrees of freedom average out when $\ell \ll L$, often resulting in a vast simplification of the equations of motion and corresponding dynamics.

Though hydrodynamics is thought of in terms of the coarse-graining of some microscopic theory, it is often the case that the microphysics of interest is not known and the fluid model is being used to make predictions in its stead.  In these cases, hydrodynamics is thought of as a long-wavelength effective field theory for a system with $\textnormal{Kn} \ll 1$.  Such a theory is typically defined by prescribing a definition for a set of conserved currents (the stress-energy tensor and a baryon current) in terms of \textit{hydrodynamic variables} drawn from equilibrium thermodynamics, common examples of which include the temperature and particle density.  This can be done by reference to the so-called \textit{gradient expansion}, which assumes (1) that the system is sufficiently ``near equilibrium'' such that it may be parameterized using variables drawn from equilibrium thermodynamics and (2) that gradients of these parameters may be treated as small quantities.  The latter assumption may be thought of as morally tantamount to the assumption that $\textnormal{Kn} \ll 1$, since the natural way to incorporate said derivative terms is through quantities of the form ${\ell \nabla \sim \ell/L \equiv \textnormal{Kn}}$.

In the gradient expansion approach, ideal (perfect fluid) hydrodynamics appears as the simplest such theory since all possible gradient terms are dropped from the definitions of the conserved currents.  In this respect, ideal hydrodynamics is \textit{zeroth order} in gradients, and perfect fluids are always in local thermodynamic equilibrium.  Effects arising outside of equilibrium (e.g. viscosity, heat conduction) are neglected at zeroth order, but can be incorporated by adding to the perfect fluid conserved currents the set of first derivative corrections, defining what is known as \textit{first-order} relativistic hydrodynamics.

In order to capture the non-equilibrium effects of viscosity and heat
conduction, Eckart \cite{Eckart_1940} and Landau \& Lifshitz \cite{Landau_1987}
independently put forth their namesake fluid theories, each of which arises at
first order in the gradient expansion.  Long before these theories could be
applied, however, they were shown to be pathological \cite{Hiscock_1983}
\cite{Hiscock_1985}, possessing acausal characteristics and unstable
equilibrium states.  At the time, these issues were incorrectly attributed to
first order hydrodynamics in general, leading to the widespread
adoption of 
an alternative formulation known as M\"uller-Israel-Stewart (MIS) theory \cite{Muller_1967} \cite{Israel:1976tn} \cite{Israel_1979} which includes both first- and second-order gradient terms\footnote{MIS-like
theories were specifically adopted to deal with the problems of Eckart
and Landau \& Lifshitz theories, 
not because second-order terms were thought
to be relevant; in fact, most studies (e.g. \cite{Saida:2010gj}
\cite{Shibata_2017} in astrophysics, and \cite{Schenke2011} \cite{Schenke_2012}
in nuclear physics) drop many of these terms to simplify computations.}.  
Though the original version of MIS theory has the same issues
with its first-order terms
(related
to the so-called hydrodynamic frame, as will be discussed later),
the problems with stability
and causality were fixed by promoting the dissipative corrections to
independent degrees of freedom complete with their own evolution equations.
This additional structure allowed for proofs of causality and stability for the
linearized theory \cite{Hiscock_1985}, and MIS-type
theories
became the standard to model relativistic dissipative fluids for decades to
come \cite{Baier:2007ix,Denicol:2012cn,Martinez:2010sc,Florkowski:2010cf}.

Inspired by a series of works due to V\'an \& Bir\'o \cite{Van:2007pw}
\cite{Van:2011yn} and Freist\"uhler \& Temple \cite{Freistuhler2014}
\cite{Freistuhler2017} \cite{Freistuhler2018}, Bemfica, Disconzi \& Noronha
\cite{Bemfica_2018} \cite{Bemfica:2020zjp} and Kovtun \cite{Kovtun_2019}
\cite{Hoult_2020} put forth a general first-order theory (which we will
refer to here as \textit{BDNK theory}) free of the pathologies of the
theories of Eckart and Landau-Lifshitz.  The key insight of these works is
that the choice of coefficients weighting the gradient terms---the
so-called \textit{hydrodynamic frame}\footnote{Here we will use the term
``frame'' to refer to hydrodynamic frame; when specifying a Lorentz
(coordinate) frame, we will use the terms ``reference frame'' or ``rest
frame.''}---must be made carefully: ``good'' choices of frame lead to
causal, stable, strongly hyperbolic theories, and ``bad'' choices (such as
those of Eckart and Landau-Lifshitz) result in unphysical ones.
The success of MIS theory in this regard derives from the ``relaxation'' form taken by the evolution equations for the dissipative degrees of freedom, which allows them to temporarily deviate from their (potentially acausal or unstable) Navier-Stokes values \cite{Romatschke_2019}. 
However, MIS-type theories
should not be viewed as
entirely distinct from the BDNK approach. In fact, it can be
shown that extensions of MIS theories to generalized hydrodynamic
frames reduce to BDNK theory instead of (acausal)
Navier-Stokes equations in the first-order limit \cite{Noronha:2021syv}. 
More work towards understanding this connection \cite{Dore:2021xqq} and in deriving fluid models from microphysical theories \cite{Rocha:2021lze} is needed to clarify this
connection.

BDNK theory possesses a number of features which may be viewed as
advantages over MIS theory.  
Most apparent of these is that BDNK theory does not possess the additional (dissipative) degrees of freedom and corresponding evolution equations which characterize MIS-type approaches.  This PDE structure
allows for rigorous proofs of
strong hyperbolicity, causality, and the stability of equilibrium states
for the class of BDNK theories where the transport coefficients satisfy a set of frame
constraints \cite{Bemfica:2020zjp}.
Similar
proofs do not exist in full generality for MIS theories, and the few that
do\footnote{MIS theory has been proven to be causal subject to a set of
  dynamical constraints, locally well-posed for Gevrey initial data when heat conduction and
  particle diffusion are neglected \cite{Bemfica_2020_MIS}, and hyperbolic
  when all dissipative effects except bulk viscosity are dropped \cite{Bemfica_2019}.} 
give constraints which are functions of both the transport coefficients
{\em and} the dissipative degrees of freedom. The latter
are ``dynamical'' in the sense that they depend on the state of the fluid,
and this is an important distinction compared to BDNK, which only has
state-independent constraints (as such
these constraints can be considered to be part
of the microphysical description of the fluid, since that determines
the transport coefficients).
For MIS this juxtaposition implies constraints have to be
monitored within a simulation to ensure they are not dynamically violated
(a step which is often omitted in the literature, leading to, for
example, a number of studies which evolved fluids in regimes
where the characteristics of the equations had superluminal speeds
\cite{Plumberg:2021bme}).  

Frame complications also cause problems for
physical scenarios relevant to astrophysics: MIS breaks down when
sufficiently strong (high Mach number) shockwaves form
\cite{Olson_1990_shocks}, whereas arbitrarily strong shockwave solutions
exist in BDNK theory for well-chosen frames \cite{Pandya_2021}
\cite{freistuhler2021nonexistence}. 
On the other hand, astrophysical problems, by and large, require the
consistent inclusion of magnetic fields into the evolution, which is
currently only known for MIS-type dissipative theories
\cite{Chandra:2015iza,Denicol:2018rbw,Denicol:2019iyh,Panda:2020zhr,Panda:2021pvq}. Despite the lack of causality
constraints, such theories have been successfully used in astrophysical
studies \cite{Chandra:2017auj,Most:2021rhr}.

Furthermore, MIS theory has the benefit of a strong base of existing numerical
infrastructure developed to model heavy-ion collisions \cite{Schenke_2012}
\cite{Takamoto_2011} \cite{Okamoto_2017}.  The existence of evolution equations
for the dissipative corrections leaves the principal part of the conservation
laws unchanged, allowing one to solve these equations in largely the same way
as in ideal hydrodynamics.
The equations used to evolve the dissipative corrections require
novel methods, however these problems were thoroughly addressed in the
heavy-ion literature mentioned previously.  BDNK theory, on the other hand, has
no additional degrees of freedom beyond ideal hydrodynamics and hence only has
the stress-energy and particle current conservation laws; however, these
equations are not amenable to the numerical methods of ideal hydrodynamics due
to the presence of first derivative terms in the conserved currents.  Thus  
novel approaches are required to apply BDNK theory in numerical models of 
physical systems.

In this study, we build upon the exploratory work of \cite{Pandya_2021} to
provide a BDNK evolution scheme with enhanced stability for flows with high
Lorentz factors, strong shockwaves, and near-vacuum states, all of which
arise generically in many physical systems of interest.  The scheme is
based on a flux-conservative finite volume discretization complete with
non-oscillatory primitive variable reconstruction and a central-upwind
numerical flux function.  We also address problems unique to the BDNK
equations, and present a novel algorithm for primitive variable recovery
capable of handling the numerically difficult inviscid limit, wherein the
BDNK equations reduce to the relativistic Euler equations.  In said limit,
our scheme reduces exactly to a high-resolution shock-capturing (HRSC)
finite volume perfect fluid solver.

We structure the remainder of the study as follows: In Sec. \ref{sec:EOM}, we begin with a brief
overview of relativistic hydrodynamics and define the BDNK conserved
currents.  To clarify our presentation, we then specialize to a fluid with
simple microphysics (namely conformal symmetry) and trivial spacetime
geometry (4D Minkowski spacetime).  In Sec. \ref{sec:numerics}, we
outline our numerical method. More precisely, we review the finite volume method, and
explain how it is applied first in ideal hydrodynamics, then how we adapt
it to a BDNK fluid.  The new code's performance is evaluated in a set of
problems with variation in one and two spatial dimensions, designed to test
constraint preservation, stability for high-velocity flows with shockwaves,
and the behavior of solutions approaching the inviscid limit; 
these tests are presented in Sec. \ref{sec:numerical_tests}. In Sec.
\ref{sec:conclusion} we conclude
with avenues for future work. In appendices we list the BDNK
conserved currents in our chosen coordinate basis, give an overview of how to
generalize the scheme beyond conformal symmetry, review the Weighted
Essentially Non-Oscillatory (WENO) algorithm, and show examples of our
scheme's convergence properties.

\section{Equations of motion} \label{sec:EOM}

Relativistic fluid models are typically defined through two conserved
currents: the stress-energy tensor $T^{ab}$ and a baryon current $J^{a}$,
each of which must obey \begin{align} \nabla_{a} T^{ab} &= 0
  \label{eq:T_cons_law}\\ \nabla_{a} J^{a} &= 0 \label{eq:J_cons_law}.
\end{align} Hydrodynamics models a substance's long-wavelength behavior by
asserting that all of the microphysical degrees of freedom average out at
the scales of interest, such that in $d$ spacetime dimensions the $d + 1$
constraints (\ref{eq:T_cons_law}-\ref{eq:J_cons_law}) are sufficient to
specify its macroscopic evolution.  These $d+1$ constraints are then
interpreted as a set of evolution equations for $d+1$ state variables,
which are typically drawn from the laws of thermodynamics, leaving the
fluid theory largely agnostic of the microphysics it is approximating.
Typical choices of these \textit{hydrodynamic variables} include the local
flow four-velocity, $u^{a}$ (assumed to be timelike), as well as a set of
scalar variables which are related to other, similar quantities by the laws
of thermodynamics.  Here we will use the energy density $\epsilon$ and the
baryon number density $n$, though these are occasionally replaced by the
temperature $T$ and chemical potential $\mu$ in the literature.

Asserting that an observer co-moving with the fluid sees energy density
$\epsilon$, an isotropic pressure $P$, and a number density $n$, gives the
\textit{perfect fluid} (\textit{ideal}) stress-energy tensor and particle
current
\begin{align}
T^{ab}_{0} &= \epsilon u^{a} u^{b} + P \Delta^{ab} \label{eq:Tab_0_cov}\\
J^{a}_{0} &= n u^{a}, \label{eq:Ja_0_cov}
\end{align}
where the tensor
\begin{equation}
\Delta^{ab} = g^{ab} + u^a u^b
\end{equation}
projects onto the space orthogonal to $u^{a}$.  Combining
(\ref{eq:Tab_0_cov}-\ref{eq:Ja_0_cov}) with
(\ref{eq:T_cons_law}-\ref{eq:J_cons_law}) yields the relativistic Euler
equations.  Note that (\ref{eq:Tab_0_cov}-\ref{eq:Ja_0_cov}) have $d+2$
parameters, rather than $d+1$ --- this implies that an additional closure
relation, $P(\epsilon, n)$, is required; the definition of $P(\epsilon, n)$ is
known as the equation of state.

One can better understand the thermodynamic properties of the perfect fluid if
one takes the projection of the relativistic Euler stress-energy conservation
equation (\ref{eq:T_cons_law}),(\ref{eq:Tab_0_cov}) along $u_b$,
\begin{equation}
\nabla_a \Big[ (\epsilon+P) u^{a} \Big] = u^a \nabla_a P.
\end{equation}
Applying the thermodynamic relation $dP = s dT + n d\mu$, where $s$ is the
entropy density,
then adding $\mu \nabla_a J_0^{a} = 0$ to the right-hand side results in
\begin{equation}
\nabla_a (\epsilon + P - \mu n) u^{a} = u^a s \nabla_a T.
\end{equation}
Using the first law of thermodynamics in intrinsic form, ${\epsilon + P - \mu n
= T s}$, one arrives at the result
\begin{equation} \label{eq:pf_entropy_cons}
\nabla_a (s u^{a}) = 0,
\end{equation}
which may be interpreted as a conservation law for the entropy current, implying
\textit{that entropy is conserved in ideal hydrodynamics}\footnote{Entropy can increase when
shocks are present, though. In these cases, the physical solution is given by
the weak formulation of the equations.  Since the weak solution may not be
unique, the physical one is that which satisfies the second law of
thermodynamics \cite{LeVeque_2006}.}.  Hence a
different theory is required to incorporate entropy-producing dissipative processes such as heat
conduction (diffusion due to thermal gradients) or viscosity (momentum transfer
due to velocity gradients).

A natural place to start when constructing a dissipative fluid theory, then, is
by including gradient terms in $T^{ab}$ and $J^{a}$.  
Formally, this procedure may be thought of as an expansion about thermodynamic equilibrium.  Since local fluctuations in the hydrodynamic variables are expected to be small, their gradients are assumed to be small as well, allowing one to stratify gradient corrections to the conserved currents by the order of derivative terms (or by the exponent of such terms, if they appear nonlinearly).
Explicitly, this
approach asserts that the ``true'' conserved currents (from the full
microphysical theory, whatever that may be) can be written in a
\textit{gradient expansion}\footnote{Though in most cases it is unknown if the gradient expansion converges, there are many known examples where it does not \cite{Heller:2013fn} \cite{Buchel:2016cbj} \cite{Denicol:2016bjh} \cite{Heller:2016rtz}.  Remarkably, in some cases even beginning with far from equilibrium and varied initial conditions, solutions still approach a similar ideal hydrodynamic evolution at late times, a phenomenon typically attributed to the existence of a universal attractor solution \cite{Heller:2015dha} \cite{Romatschke:2017vte}.} 
\begin{equation}\label{eq:gradient_exp}
\begin{aligned}
T^{ab} &= T^{ab}_{0} + O(\nabla) + O(\nabla^2) + ... \\
J^{a} &= J^{a}_{0} + O(\nabla) + O(\nabla^2) + ... \\
\end{aligned}
\end{equation}
where the $O(\nabla)$ terms include only first gradients, the $O(\nabla^2)$ terms include second gradients and products of first gradients, and so on.  In practice, it is impossible to construct $T^{ab}, J^{a}$ up to infinite order in derivative terms, so one typically truncates them after including all such terms up to a given order\footnote{Though one can certainly write down higher-order gradient theories (see, e.g., the third-order theory of \cite{Grozdanov_2016}), the gradient expansion written in (\ref{eq:trunc_grad_exp}) is not meant to imply that higher-order theories are always strictly ``better'' than lower-order ones.  In fact, it has been shown that inconsistencies arise already at second order in gradients, both in nonrelativistic contexts (super-Burnett theory) \cite{DESCHEPPER19741} and relativistic contexts (MIS theory) \cite{Kovtun:2011np}.} 
$n$, denoted here with a subscript:
\begin{equation}\label{eq:trunc_grad_exp}
\begin{aligned}
T^{ab}_{n} &= T^{ab}_{0} + O(\nabla) + ... + O(\nabla^{n}) \\
J^{a}_{n} &= J^{a}_{0} + O(\nabla) + ... + O(\nabla^{n}).
\end{aligned}
\end{equation}
By construction, the perfect fluid stress-energy tensor and particle current are
recovered when all derivative corrections to $T^{ab}, J^{a}$ are dropped; in
other words, the perfect fluid arises from truncating the gradient expansion at
zeroth order, keeping only $T^{ab}_0, J^{a}_0$.

BDNK theory \cite{Bemfica:2020zjp} arises when truncating the gradient
expansion at first order\footnote{The aforementioned MIS theory may be interpreted as arising from a truncation at second order in gradients, though it was not originally derived by reference to a gradient expansion; see \cite{Pandya_2021} \cite{Baier:2007ix}.}, and defines
$T^{ab}_1, J^{a}_1$ by taking linear combinations of all allowed one-derivative
terms, weighted by zero-derivative transport coefficients.  In
\cite{Bemfica:2020zjp}, the authors also provide a set of conditions on these
transport coefficients which guarantee the theory be strongly hyperbolic,
causal, consistent with the second law of thermodynamics within the regime of
validity of the gradient expansion (e.g. $\nabla_a (s u^a) \geq 0 +
O(\nabla^3)$, cf. (\ref{eq:pf_entropy_cons})) and have stable equilibrium
states.

The BDNK conserved currents are
\begin{multline} \label{eq:Tab_cov}
T^{a b}_{1} = (\epsilon + \mathcal{A}) u^a u^b + (P + \Pi) \Delta^{ab} + \mathcal{Q}^a u^b + \mathcal{Q}^b u^a \\
- 2 \eta \sigma^{ab}
\end{multline}
and
\begin{equation} \label{eq:J_cov}
J^{a}_{1} = \mathcal{N} u^{a} + \mathcal{J}^{a},
\end{equation}
where each has dissipative contributions linear in gradients of the hydrodynamic variables.  These gradient corrections to $T^{ab}$ are defined to be
\begin{align}
\mathcal{A} &= \tau_{\epsilon} \big[ u^c \nabla_c \epsilon + (\epsilon + P) \nabla_c u^c \big] \label{eq:A_cov}\\
\Pi &= -\zeta \nabla_c u^c + \tau_{P} \big[ u^c \nabla_c \epsilon + (\epsilon + P) \nabla_c u^c \big] \\
\mathcal{Q}^{a} &= \tau_{Q} (\epsilon + P) u^c \nabla_c u^a + \beta_{\epsilon} \Delta^{a c} \nabla_c \epsilon + \beta_{n} \Delta^{a c} \nabla_c n
\end{align}
and
\begin{multline} \label{eq:shear_cov}
\sigma^{ab} = \frac{1}{2} \Big( \Delta^{ac} \Delta^{bd} \nabla_c u_d + \Delta^{ac} \Delta^{bd} \nabla_d u_c \\
- \frac{2}{3} \Delta^{ab} \Delta^{cd} \nabla_c u_d \Big),
\end{multline}
which are the correction to the energy density, the bulk viscous pressure, the
heat flow vector, and the shear tensor, respectively.  
It turns out that one may drop all gradient corrections to the particle current without compromising the hyperbolicity, causality, or thermodynamic stability properties of the resulting PDEs \cite{Bemfica:2020zjp}.  We make such a choice here and set
\begin{equation} \label{eq:J_grad_correction}
\mathcal{N} = n, ~~~ \mathcal{J}^{a} = 0.
\end{equation}
Note that the particle
current (\ref{eq:J_cov},\ref{eq:J_grad_correction}) now takes the same form as in ideal hydrodynamics, and
that one recovers the perfect fluid stress-energy tensor (\ref{eq:Tab_0_cov}) from (\ref{eq:Tab_cov}) if the gradient corrections $\mathcal{A}, \Pi, \mathcal{Q}^{a}, \eta \sigma^{ab}$ all vanish.

Each of the gradient terms is linear in one of the transport
coefficients, which themselves are free of derivatives and are derived from
the thermodynamics of the specific substance being modeled.  Inspection of the
terms above shows that these coefficients fall into three categories. The first
are thermal transport coefficients,
\begin{equation}
\begin{aligned}
\beta_{\epsilon} &= \tau_{Q} \Big(\frac{\partial P}{\partial \epsilon}\Big)_{n} + \frac{\sigma T (\epsilon + P)}{n} \Big( \frac{\partial (\mu/T)}{\partial \epsilon} \Big)_{n} \\
\beta_{n} &= \tau_{Q} \Big( \frac{\partial P}{\partial n} \Big)_{\epsilon} + \frac{\sigma T (\epsilon + P)}{n} \Big( \frac{\partial (\mu/T)}{\partial n} \Big)_{\epsilon},
\end{aligned}
\end{equation}
which depend on derivatives of the equation of state $P(\epsilon, n)$ and
chemical potential divided by the temperature, $\frac{\mu}{T}(\epsilon, n)$,
which must be computed via the laws of thermodynamics.  Next are the transport
coefficients corresponding to well-known dissipative effects, namely the shear
viscosity $\eta$, bulk viscosity $\zeta$, and thermal conductivity $\sigma$
(which appears in $\beta_{\epsilon}, \beta_{n}$).  Finally there are a set of
three relaxation times $\tau_{\epsilon}, \tau_{Q}, \tau_{P}$ which set the
dissipative timescales.

Here, as in \cite{Pandya_2021}, for the sake of simplicity we specialize to a
fluid with an underlying conformal symmetry (which requires $g_{ab} T^{ab} = 0$) and we set $\mu = 0$.  Together
these conditions\footnote{A conformal fluid with $\mu = 0$ is often used as a simple toy model for the QGP produced in heavy ion collisions; for more on conformal fluids, see \cite{Baier:2007ix}.} imply
\begin{equation}
\begin{aligned}
P(\epsilon, n) = \frac{\epsilon}{3}, ~~~~~ \Pi &= \frac{\mathcal{A}}{3}, ~~~~~ \zeta = 0, \\
\beta_{\epsilon} = \frac{\tau_{Q}}{3}, ~~~~~ \beta_{n} &= 0, ~~~~~ \tau_{P} = \frac{\tau_{\epsilon}}{3},
\end{aligned}
\end{equation}
leaving us with only the shear viscosity $\eta$ and the two relaxation times
$\tau_{\epsilon}, \tau_{Q}$.  Note that the stress-energy tensor is now free of
$n$, and hence $J^{a}_1$ (\ref{eq:J_cov}) and $T^{ab}_1$ (\ref{eq:Tab_cov})
decouple.  In \cite{Pandya_2021}, this fact is used to neglect the evolution of
the particle current; here we choose to evolve $J_1^{a}$ so that we may use the
number density $n$ as an intuitive marker of the fluid's behavior in the
tests below.

For the remaining three transport coefficients we follow the same prescription
as in \cite{Pandya_2021}, adopting natural units and writing them as
\begin{equation} \label{eq:frame_coeffs_def}
\begin{aligned}
\eta &\equiv \eta_0 \epsilon^{3/4} \\
\tau_{\epsilon} = \frac{3}{4 \epsilon} \chi &\equiv \frac{3}{4 \epsilon} \chi_0 \epsilon^{3/4} \\
\tau_{Q} = \frac{3}{4 \epsilon} \lambda &\equiv \frac{3}{4 \epsilon} \lambda_0 \epsilon^{3/4} \\
\end{aligned}
\end{equation}
where $\tau_{\epsilon}, \tau_{Q}$ are exchanged for $\chi, \lambda$ to parallel
the notation of \cite{Bemfica_2018}, and then are written with the $\epsilon$
dependence pulled out.  Writing them in this way allows us to use the
dimensionful constant $\eta_0$ as a free parameter controlling the
amount of viscosity in the model, and the remaining two constants $(\chi_0,
\lambda_0)$ determine the so-called ``hydrodynamic frame.''  
In accordance with \cite{Pandya_2021} we choose the frame
\begin{equation} \label{eq:frame_coeffs}
(\chi_0, \lambda_0) = \Big( \frac{25}{4} \eta_0, \frac{25}{7} \eta_0 \Big),
\end{equation}
which fixes the characteristic speeds to be exactly unity.
This choice is consistent with the
conditions of \cite{Bemfica_2018} which establish existence and uniqueness of
solutions, causality, and linear stability about thermodynamic equilibrium,
and those of \cite{freistuhler2021nonexistence} establishing the existence
of smooth strong shock solutions.

With the transport coefficients written in the form (\ref{eq:frame_coeffs}),
the limit $\eta_0 \to 0$ results in $\chi_0, \lambda_0 \to 0$, and all
dissipative corrections in the BDNK stress-energy tensor
(\ref{eq:A_cov}-\ref{eq:shear_cov}) vanish, reducing the BDNK conserved
currents exactly to those of the perfect fluid
(\ref{eq:Tab_0_cov}-\ref{eq:Ja_0_cov}).  We refer to
$\eta_0 \to 0$ as the \textit{inviscid limit} of BDNK theory.

In the work that follows, we further specialize to a fluid in 4D Minkowski
spacetime and adopt a Cartesian coordinate system $x^{a} = (t, x, y, z)^{T}$.
To limit computational cost, we only consider test problems with variation in
one spatial dimension $(t, x)$ or two spatial dimensions $(t, x, y)$.  The
components of $T_1^{ab}, J_1^{a}$ in these coordinates are rather long, so we
relegate them to Appendix \ref{sec:coord_eqns}.

In the following section we review the finite volume method and how it is
typically applied to the perfect fluid equations of motion
(\ref{eq:T_cons_law}-\ref{eq:J_cons_law},
\ref{eq:Tab_0_cov}-\ref{eq:Ja_0_cov}), before adapting it to the BDNK equations
(\ref{eq:T_cons_law}-\ref{eq:J_cons_law}, \ref{eq:Tab_cov}-\ref{eq:J_grad_correction}).

\section{Numerical algorithm} \label{sec:numerics}

In this section we outline the finite volume method, then describe how one casts the relativistic Euler and BDNK equations into the conservative form required for its application.  We then detail the steps in the finite volume algorithm as they are applied to the BDNK equations in 4D Minkowski spacetime, assuming one is only interested in problems with variation in two spatial dimensions (though the methods straightforwardly generalize to higher dimensional problems).  

In Sec. \ref{sec:pvr} we discuss primitive variable recovery, which is trivial for BDNK theory, as the BDNK stress-energy tensor is linear in its primitive variables.  Care is required to apply this analytic solution for small viscosities, however, as the solution breaks down in cells where the viscous terms are 
unresolved (smaller in magnitude than truncation error). 
We detail an adaptive algorithm which applies the perfect fluid's primitive variable solver in unresolved cells, allowing for a stable evolution at such ``low'' resolutions. 

Sec. \ref{sec:reconstruction} outlines the reconstruction of primitive variables, for which we use the WENO method \cite{Shu1998}. For BDNK one must also compute spatial derivatives of primitive variables prior to reconstruction, and for that we use a method based on the central-WENO approach \cite{Levy1999}.  For the numerical fluxes, we use the Kurganov-Tadmor flux function \cite{Kurganov_2000} and set the maximum local speed $a = 1$, which is the exact local characteristic speed for the BDNK equations in the chosen hydrodynamic frame (\ref{eq:frame_coeffs}).  

We conclude this section with a brief discussion of the time integration algorithm used (Heun's method) and comment on the choice of boundary conditions.

\subsection{Finite volume method} \label{sec:fvm}

Though it is not yet clear whether the BDNK equations possess sensible discontinuous
shockwave solutions, it is well known that such solutions are essential in modeling
perfect fluids, which arise in the inviscid limit of BDNK theory.
As a result, it would be preferable to develop a numerical method for the BDNK
equations which can stably evolve solutions with sharp gradients.  For this
reason, we adopt a \textit{finite volume} discretization of the BDNK PDEs.
Before doing so, however, we will first briefly review finite volume methods in
the context of relativistic fluid mechanics.

Finite volume methods are adapted to the solution of hyperbolic conservation
laws, which in general may be written in the form
\begin{equation}\label{eq:conserv_law}
\frac{\partial}{\partial t} \bm{q}(\bm{p}) + \frac{\partial}{\partial x^{i}} \bm{f}^{i}(\bm{p}) = \bm{\psi}(\bm{p}),
\end{equation}
where $\bm{q}$ is a vector of \textit{conservative variables}, $\bm{f}$ is the
\textit{flux} tensor, $\bm{\psi}$ is a vector of \textit{sources}, and each of
the aforementioned terms is a function of the vector of \textit{primitive
variables} $\bm{p}$.

Assuming one wants to solve the conservation law (\ref{eq:conserv_law}) over a
spatial domain $\mathcal{D}$ (which we will take to be two-dimensional), one
divides $\mathcal{D}$ into subdomains $\mathcal{S}_{i,j}$, which we will define
to be rectangular with extent $|\mathcal{S}_{i,j}| = \Delta x \Delta y$,
centered at the point $(x, y)$.  One may integrate
(\ref{eq:conserv_law}) inside a given subdomain to get
\begin{multline} \label{eq:finite_vol}
\frac{\partial}{\partial t} \bar{\bm{q}}_{i,j} + \frac{\langle \bm{f}^{x}_{i+\frac{1}{2},j} \rangle - \langle \bm{f}^{x}_{i - \frac{1}{2},j} \rangle}{\Delta x} \\
+ \frac{\langle \bm{f}^{y}_{i,j+\frac{1}{2}} \rangle - \langle \bm{f}^{y}_{i,j - \frac{1}{2}} \rangle}{\Delta y} = \bar{\bm{\psi}}_{i,j},
\end{multline}
where we have introduced the shorthand
\begin{align}
\bar{X}_{i} &= \frac{1}{\Delta x \Delta y} \int_{\mathcal{S}_{i,j}} X \, dx dy \label{eq:cell_avg}\\
\langle X^{k} \rangle &= \int_{\partial \mathcal{S}_{i,j}} X^{k} \, dx^{l}, ~~~ k \neq l, \label{eq:face_avg}
\end{align}
so an overbar denotes a cell-averaged quantity, and angle brackets
denote a flux in direction $k$ averaged over the face of the cell at constant coordinate $x^{k}$.

As written, (\ref{eq:finite_vol}) may be interpreted as a semidiscrete
evolution equation for the cell-averages $\bar{\bm{q}}_{i,j}$ after choosing
suitable discrete approximation to the integrals
(\ref{eq:cell_avg}-\ref{eq:face_avg}) and reinterpreting all variables as
discrete grid functions defined on the cells $\mathcal{S}_{i,j}$.

By discretizing the integral form of the conservation law (\ref{eq:finite_vol})
rather than the PDE (\ref{eq:conserv_law}), the finite volume method enjoys a
number of advantages over other methods (such as the finite difference
method\footnote{It is also possible to construct a conservative finite
difference scheme, provided one defines the flux derivative term by
reference to integrals of the flux; see \cite{Shu1998} \cite{Rezzolla2013}.}
used in \cite{Pandya_2021}).  Most important of these for our purposes is the
ability to stably evolve shockwave solutions, which are generically
discontinuous for inviscid flows.  Such solutions satisfy the weak (integral)
form of the equations (\ref{eq:finite_vol}) but not the continuum PDE
(\ref{eq:conserv_law}), and hence may be recovered by virtue of discretizing
(\ref{eq:finite_vol}) rather than (\ref{eq:conserv_law}).  It is important to
note, though, that the precise choice of discretization for the numerical
fluxes $\bm{F}$ approximating $\langle \bm{f} \rangle$ has a significant impact
on stability as well as constraint preservation (namely conservation of
$\bm{q}$, modulo sources $\bm{\psi}$, across the simulation domain and preservation of the
irrotational nature of the spatial gradients $\partial_i \bm{p}$); these topics are discussed further
in Sec. \ref{sec:numerical_flux}.

Since the conservation law (\ref{eq:finite_vol}) is discretized over a
finite-sized cell (rather than, say, a zero-volume point, as is the case for
the finite difference method), there are a number of additional considerations
which appear when solving the discrete equations.  Consider taking a time step
of the latter beginning at time
level index $n$ (either from initial data specified then, or after a prior successful time step); at this point, the
cell-averaged conservative variables $\bar{\bm{q}}^{n}_{i,j}$ are known at time
level $n$ for all of the spatial grid points indexed with $i,j$.  To
use the discrete version of (\ref{eq:finite_vol}) to find the conservative
variables at the next time level, $\bar{\bm{q}}^{n+1}_{i,j}$, one has to take the
following steps: 
\begin{enumerate}
    \item To solve (\ref{eq:finite_vol}), one needs to compute the flux terms
    $\langle \bm{f} \rangle$ and the source term $\bar{\bm{\psi}}$, which are
    functions of the primitive variables $\bm{p}$.  This is done by inverting
    the definitions of the (known) conservative variables, $\bm{q}(\bm{p})$, to
    find $\bar{\bm{p}}^{n}_{i,j}$.  This step is known as \textit{primitive
    variable recovery}.
    \item Once the primitive variables are known, the source term
    $\bar{\bm{\psi}}^{n}_{i,j}$ can be trivially computed.  Computing the flux
    terms is not so straightforward, however, since these are averaged over
    cell faces (\ref{eq:face_avg}) and the primitive variables we have computed
    are cell-averages $\bar{\bm{p}}^{n}_{i,j}$.  Hence one must interpolate the
    primitive variables from the cell-average $\bar{\bm{p}}^{n}_{i,j}$ to the
    interfaces $\langle \bm{p}^{n}_{i,j} \rangle$ in a step known as
    \textit{reconstruction}.
    \item Using the interface-averaged primitive variables $\langle
    \bm{p}^{n}_{i,j} \rangle$ one can finally \textit{compute the numerical
    fluxes} approximating $\langle \bm{f} \rangle$.  The discretization
    (\ref{eq:finite_vol}) may now be solved for the cell-averaged conservative
    variables at the next time level, $\bar{\bm{q}}^{n+1}_{i,j}$.
\end{enumerate}

In the following sections we explain in detail how each of the above steps is
carried out, first for the relativistic Euler equations, and then for the BDNK
equations.  We begin by casting both sets of equations into conservative form
(\ref{eq:conserv_law}), then address primitive variable recovery,
reconstruction, and numerical flux computation in successive subsections.  We
conclude the section with a brief discussion of the time integration algorithm
and the types of boundary conditions implemented for the numerical tests which
follow.

\subsection{Relativistic fluid equations in conservative form}

Both the relativistic Euler equations (\ref{eq:T_cons_law}-\ref{eq:Ja_0_cov})
and the BDNK equations
(\ref{eq:T_cons_law}-\ref{eq:J_cons_law},\ref{eq:Tab_cov}-\ref{eq:J_grad_correction})
can be cast into the form (\ref{eq:conserv_law}) in the same way.  Combining
the different components of the equation into vectors, one can write
(\ref{eq:T_cons_law}) as
\begin{equation} \label{eq:delT_conserv_form}
\bm{q} =
\begin{pmatrix}
T^{tt} \\
T^{tx} \\
T^{ty} \\
\end{pmatrix}
,~~
\bm{f}^{x} = 
\begin{pmatrix}
T^{tx} \\
T^{xx} \\
T^{yx} \\
\end{pmatrix}
,~~
\bm{f}^{y} = 
\begin{pmatrix}
T^{ty} \\
T^{xy} \\
T^{yy} \\
\end{pmatrix}
,~~
\bm{\psi} = \bm{0},
\end{equation}
where each equation comes from a row of the vectors above.  For example, the
first equation is ${T^{tt}_{,t} + T^{tx}_{,x} + T^{ty}_{,y} = 0}$.  

The particle current conservation law (\ref{eq:J_cons_law}) is a scalar
equation, and may be written
\begin{equation} \label{eq:delJ_conserv_form}
q = J^{t}, ~~ f^{x} = J^{x}, ~~ f^{y} = J^{y}, ~~ \psi = 0;
\end{equation}
as mentioned before, the particle current is identical between the relativistic
Euler and BDNK equations.
As a result, one may evolve the particle current (and hence $n$) forward through time using standard methods used to solve the equations of ideal hydrodynamics.  In the sections that follow, we focus on the methods used to solve (\ref{eq:T_cons_law}).  After defining these methods, we briefly summarize how they are applied to solve (\ref{eq:J_cons_law}) in Sec. \ref{sec:evolving_J}.

Though (\ref{eq:delT_conserv_form}) appears to be essentially
identical between the zeroth and first-order theories, differences arise in the
primitive variable recovery step (because each has a different set of primitive
variables), in reconstruction, as well as in the computation of the flux terms.
These differences will be described in the following three subsections.

\subsection{Primitive variable recovery} \label{sec:pvr}
For a conformal fluid in 4D Minkowski spacetime with variation in $(t, x, y)$,
the set of primitive variables for the perfect fluid
(\ref{eq:Tab_0_cov}-\ref{eq:Ja_0_cov}) are
\begin{equation}\label{eq:PF_p0}
\bm{p}_{0} = 
\begin{pmatrix}
\epsilon \\
u^{x} \\
u^{y}
\end{pmatrix}.
\end{equation}
The primitive variable solution
$\bm{p}_{0}(\bm{q}_{0})$ can be carried out analytically in this case, and is
given by
\begin{equation} \label{eq:PF_pvr}
\begin{aligned}
\epsilon &= - T^{tt} + \sqrt{ 6 (T^{tt})^2 + 3 [(T^{tt})^2 - (T^{tx})^2 - (T^{ty})^2]} \\
|v| &= \frac{\sqrt{(T^{tx})^2 + (T^{ty})^2}}{T^{tt} + 3 \epsilon}, ~~~ 
u^{t} = \frac{1}{\sqrt{1 - |v|^2}} \\
u^{x} &= \frac{3 u^{t} T^{tx}}{3 T^{tt} + \epsilon}, ~~~
u^{y} = \frac{3 u^{t} T^{ty}}{3 T^{tt} + \epsilon}.
\end{aligned}
\end{equation}
It is important to stress that in general, the primitive variable solution
analogous to (\ref{eq:PF_pvr}) cannot be found analytically; the fact that it
can be here is a result of conformal symmetry, the choice of Cartesian
coordinates, and the flat spacetime background.

To write the BDNK equations---which, unlike the relativistic Euler equations, are second-order PDEs---in
conservative form, one must perform a first-order reduction, defining the BDNK
primitive variables in terms of time derivatives of the hydrodynamic variables.
Explicitly, one such choice would be to take $\bm{p}_1 = \dot{\bm{p}}_0 =
(\dot{\epsilon}, \dot{u}^{x}, \dot{u}^{y})$, where an overdot is shorthand for
a time derivative, $\dot{X} \equiv \partial_t X$.  Here, for improved
stability\footnote{We find that the primitive variables of \cite{Pandya_2021},
$\epsilon \in (0, \infty), v^{i} \in (-1, 1)$ can reach unphysical values as a
result of numerical error in the primitive variable recovery step.  To avoid
this problem, we instead evolve $\xi \equiv \ln(\epsilon)$ and $u^{i}$, whose
values are physical for $\xi, u^{i} \in (-\infty, \infty)$.} we evolve $\xi
\equiv \ln(\epsilon)$ instead of $\epsilon$, and hence we take the BDNK
primitive variables to be
\begin{equation} \label{eq:BDNK_p1}
\bm{p}_1 =
\begin{pmatrix}
\dot{\xi} \\
\dot{u}^{x} \\
\dot{u}^{y}
\end{pmatrix}.
\end{equation}
Performing the first order reduction implies that the system
(\ref{eq:conserv_law}), (\ref{eq:delT_conserv_form}) must be augmented with a
set of trivial evolution equations used to update the hydrodynamic variables
given their time derivatives; in this case, these equations are
\begin{equation} \label{eq:trivial_evol_eqns}
\frac{\partial \xi}{\partial t} = \dot{\xi}, ~~~~~ \frac{\partial u^{x}}{\partial t} = \dot{u}^{x}, ~~~~~ \frac{\partial u^{y}}{\partial t} = \dot{u}^{y}.
\end{equation}

Brief inspection of (\ref{eq:Tab_cov}) would seem to imply that the primitive
variable recovery would be very difficult for the BDNK equations, as the
definition of the stress-energy tensor is much more complicated than it is in
the perfect fluid case (\ref{eq:Tab_0_cov}), where the primitive variable
solution is generally impossible to perform analytically.  It turns out,
however, that since $T^{ab}_1$ is linear in gradient terms by construction,
(\ref{eq:Tab_cov}) is actually of the form
\begin{equation}\label{eq:BDNK_linearity_in_p}
\bm{q}_1 = \bm{q}_0(\bm{p}_0) + \eta_0 \Big[ \bm{A}(\bm{p}_0) \cdot \bm{p}_1 + \bm{b}(\bm{p}_0, \partial_i \bm{p}_0) \Big],
\end{equation}
where we will use uppercase bold letters to denote matrices and lowercase bold 
letters for vectors.  Written in the form (\ref{eq:BDNK_linearity_in_p}), it is
clear that 
\begin{equation} \label{eq:naive_BDNK_pvr}
\bm{p}_1 = \bm{A}^{-1} \cdot \Big[ \frac{1}{\eta_0} (\bm{q}_1 - \bm{q}_0) - \bm{b} \Big],
\end{equation}
so the BDNK primitive variable solution can always\footnote{The primitive variable solution (\ref{eq:naive_BDNK_pvr}) requires $\bm{A}^{-1}$ to exist, which is always the case for physical values of the hydrodynamic variables in the chosen hydrodynamic frame.} 
be obtained analytically.  In
this sense, primitive variable recovery is actually \textit{simpler} for BDNK
than it is for the relativistic Euler equations.

Though it is straightforward to derive the BDNK primitive variable solution, (\ref{eq:naive_BDNK_pvr}) cannot be naively applied in all cases of interest. In particular, the limit $\eta_0 \to 0$ causes significant problems in numerical simulations, where truncation error $\bm{\tau}$ is introduced and (\ref{eq:BDNK_linearity_in_p}) becomes
\begin{equation}\label{eq:numerical_BDNK_q_1}
\bm{q}_1 = \bm{q}_0(\bm{p}_0) + \eta_0 \Big[ \bm{A}(\bm{p}_0) \cdot \bm{p}_1 + \bm{b}(\bm{p}_0, \partial_i \bm{p}_0) \Big] + \bm{\tau}.
\end{equation}
Note that truncation error appears as an additional correction to $\bm{q}_0$, much like the viscous term proportional to $\eta_0$; in this sense, $\bm{\tau}$ may be thought of as the contribution of \textit{numerical viscosity} to the solution.  Solving for $\bm{p}_1$ becomes difficult in cases where $\eta_0$ is so small that ${\eta_0 [\bm{A}\cdot \bm{p}_1 + \bm{b}] \lesssim \bm{\tau}}$, as (\ref{eq:numerical_BDNK_q_1}) effectively becomes inviscid up to truncation error,
\begin{equation} \label{eq:unresolved_visc}
\bm{q}_1 \approx \bm{q}_0(\bm{p}_0) + \bm{\tau},
\end{equation}
and naive application of (\ref{eq:naive_BDNK_pvr}) yields
\begin{equation}
\bm{p}_1 \approx \bm{A}^{-1} \cdot \Big[ \frac{\bm{\tau}}{\eta_0} - \bm{b} \Big],
\end{equation}
where the first term is numerical error amplified by the large factor $\eta_0^{-1}$.  This problem may be stated succinctly as follows: the BDNK primitive variable solution (\ref{eq:naive_BDNK_pvr}) \textit{breaks down whenever the numerical viscosity is comparable to or larger than the physical viscosity.}

In principle, one may be interested in solving the BDNK equations for arbitrarily small viscosities at finite grid resolution.  Here we present an adaptive algorithm to handle such cases, where cells in which the physical viscosity is unresolved (cf. (\ref{eq:unresolved_visc})) use the perfect fluid primitive variable solution, and those where it is resolved use a variant of (\ref{eq:naive_BDNK_pvr}).  The criterion used to designate a cell as viscous or inviscid preferentially uses the former as resolution is increased, eventually using the viscous solution exclusively at sufficiently high resolution.  This process should provide stable results at low resolution which converge to solutions of the continuum BDNK PDEs as the grid is refined.

To develop this adaptive scheme, we begin by examining the expected behavior of $\bm{p}_1$ at $\eta_0 = 0$, where (\ref{eq:naive_BDNK_pvr}) is indeterminate.  When $\eta_0 = 0$, the time derivative terms $\bm{p}_1$ do not appear in the conservative variables, but instead in the equations of motion, which are linear in said terms and may be written (in non-conservative form) as
\begin{equation} \label{eq:RE_eqns_nonconservative}
\bm{p}_1^{PF} = \bm{c}(\bm{p}_0, \partial_{i} \bm{p}_0),
\end{equation}
where the superscript $PF$ has been appended to denote that these variables are computed using the perfect fluid equations of motion.  Ideally, one would want (\ref{eq:naive_BDNK_pvr}) to give $\bm{p}_1 \to \bm{p}_1^{PF}$ as $\eta_0 \to 0$; this can be done in practice by defining a new set of variables\footnote{Note that $\tilde{\bm{q}}_1$ is not evolved; the standard conservative variables $\bm{q}_1$ are evolved via (\ref{eq:T_cons_law}-\ref{eq:J_cons_law}), and the shifted variables $\tilde{\bm{q}}_1$ are computed from $\bm{q}_1$ via (\ref{eq:q_tilde_defn}) during the primitive variable recovery step.},
\begin{equation} \label{eq:q_tilde_defn}
\tilde{\bm{q}}_1 \equiv \bm{q}_1 - \bm{q}_1 \Big|_{\bm{p}_1 \to \bm{p}_1^{PF}} = \eta_0 \bm{A} \cdot (\bm{p}_1 - \bm{p}_1^{PF}),
\end{equation}
where the second equality comes from applying (\ref{eq:BDNK_linearity_in_p}). Inverting
$\tilde{\bm{q}}_1(\bm{p}_1)$ yields
\begin{equation} \label{eq:shifted_BDNK_pvr}
\bm{p}_1 = \frac{1}{\eta_0} \bm{A}^{-1} \cdot \tilde{\bm{q}}_1 + \bm{p}_1^{PF}.
\end{equation}
As written, (\ref{eq:shifted_BDNK_pvr}) suffers from the same problem as (\ref{eq:naive_BDNK_pvr})---truncation error appearing in $\bm{A}^{-1}\cdot \tilde{\bm{q}}_1$ destabilizes the scheme when $\eta_0$ is sufficiently small.  To address this issue, we use (\ref{eq:shifted_BDNK_pvr}) in the following algorithm:
\begin{enumerate}
\item Compute an estimate for the numerical viscosity, which we use to define the ``viscous tolerance'' $\Delta_\eta$.
\item Compute $\tilde{\bm{q}}_1$ using (\ref{eq:q_tilde_defn}).
\item Compare $\tilde{\bm{q}}_1$ to $\Delta_\eta$:
\begin{enumerate}
\item if $\tilde{\bm{q}}_1 \geq \Delta_\eta$, use (\ref{eq:shifted_BDNK_pvr}) as-is to find $\bm{p}_1$.  Update $\bm{p}_0$ terms using the trivial evolution equations, (\ref{eq:trivial_evol_eqns}).
\item if $\tilde{\bm{q}}_1 < \Delta_\eta$, use
(\ref{eq:shifted_BDNK_pvr}) with $\tilde{\bm{q}}_1 = 0$ to
compute $\bm{p}_1$. Since the conservation law
(\ref{eq:T_cons_law}) decouples from (\ref{eq:shifted_BDNK_pvr})
when $\tilde{\bm{q}}_1 = 0$, one must update $\bm{p}_0$ using the
perfect fluid primitive variable solution (\ref{eq:PF_pvr}).
As a consequence, in this case \eqref{eq:trivial_evol_eqns} is no
longer used.
\end{enumerate}
\end{enumerate} 
As explained above, this algorithm is able to construct convergent solutions for arbitrarily small $\eta_0$ as long as the viscous tolerance $\Delta_\eta$ is lowered as the resolution is increased.  Ideally, one would compute $\Delta_\eta$ using a method to estimate the local truncation error in the cell, perhaps using an approach based in Richardson extrapolation as is done in adaptive mesh refinement schemes \cite{Berger1984}; here we adopt a simple empirical approach, tuning $\Delta_\eta$ on a problem-by-problem basis to be as small as possible without compromising the stability of the numerical solution.  Tests illustrating the behavior and convergence properties of the scheme in the $\eta_0 \to 0$ limit are shown in Sec. \ref{sec:results_inv_limit}.

Though this section is specialized to primitive variable recovery for a conformal BDNK fluid, it generalizes to non-conformal fluids in a straightforward way---see Appendix \ref{sec:non_conformal_pvr}.

\subsection{Reconstruction} \label{sec:reconstruction}

As can be seen from (\ref{eq:delT_conserv_form}-\ref{eq:delJ_conserv_form}),
both the relativistic Euler and BDNK equations have fluxes which take roughly
the same form.  Both include the terms $\bm{p}_0 = (\epsilon, u^x, u^y)^{T}$,
which must be reconstructed at the cell interfaces from their cell-averaged
values $\bar{\bm{p}}_{0}$ after primitive variable recovery.  Though there are
many different reconstruction algorithms (see \cite{LeVeque_2006}
\cite{Font_2000} for a review), we use the fifth-order Weighted Essentially
Non-Oscillatory method, WENO \cite{Liu1994} \cite{Jiang1996}.
We provide a review of WENO reconstruction in Appendix \ref{sec:WENO}.  For the sake of simplicity, for the remainder of this section and in Appendix \ref{sec:WENO} we specialize to problems with variation in 1D, as the methods described generalize to higher dimensions by simple repeated application of the 1D algorithms.

The WENO procedure mentioned above may be used to reconstruct all of the
variables present in the perfect fluid fluxes, $\bm{p}_0$.  The same cannot be
said for the BDNK fluxes, however, as they also include spatial derivative
terms proportional to $\partial_{i} \bm{p}_{0}$ such as, e.g. $u^{x}_{,x}$;
prior to reconstructing the values of these terms at the interfaces, one must
first compute the needed derivatives.  For smooth flows, it suffices to use standard
finite difference stencils to compute the derivative terms.  For flows with
sharp gradients, however, these finite differences result in the formation of
spurious oscillations, which in turn produce unphysical fluid states that destabilize the primitive variable recovery step (\ref{eq:shifted_BDNK_pvr}).  
To mitigate this instability, we instead compute
the derivative terms using an adaptive procedure based in the central-WENO
(CWENO) method of \cite{Levy1999}, whereby three different candidate stencils
are combined to minimize spurious oscillations near sharp gradients.

To achieve this non-oscillatory property, CWENO produces an interpolation
polynomial using a nonlinear weighted sum of ENO polynomials of the
cell-averages $\bar{p}_{i}$ as in WENO.  Unlike WENO, however, CWENO uses ENO
polynomial stencils which are centered about the interface rather than being
left- or right-biased.  To
apply CWENO to compute derivatives, we take the CWENO interpolation polynomial
$p_j(x^{i})$ and we differentiate it with respect to $x$ to get $p'_j(x^{i})$.
We can then evaluate this polynomial at the center of the cell of interest,
which yields
\begin{multline} \label{eq:CWENO_derivative}
\bar{p}'_{i} = \frac{\bar{p}_{i-2}-4 \bar{p}_{i-1} + 3 \bar{p}_{i}}{2 h} w_{0} + \frac{\bar{p}_{i+1}-\bar{p}_{i-1}}{2 h} w_{1} \\
+ \frac{-3 \bar{p}_{i} + 4 \bar{p}_{i+1} - \bar{p}_{i+2}}{2 h} w_{2},
\end{multline}
which is a weighted sum of the second-order backward, centered, and forward
finite difference stencils for a first derivative in $x$, where $h$ is the grid spacing.
The nonlinear weights $w_k$, (\ref{eq:WENO_weights}), are defined the same way as in the WENO case with the same smoothness indicators, except the corresponding linear weights (which appear in (\ref{eq:WENO_weights})) are modified to be
\begin{equation}
d_k = \Big( \frac{1}{6}, \frac{2}{3}, \frac{1}{6} \Big)
\end{equation}
and give fourth-order accuracy in the derivative (\ref{eq:CWENO_derivative}).  

Both the WENO reconstruction and CWENO derivative computation
depend on a free parameter $\epsilon_{W}$ (\ref{eq:WENO_weights})
controlling the amount of sensitivity each step has to sharp gradients in one of the candidate stencils. 
In principle we can have different values for $\epsilon_{W}$ in these two steps, either
to make the reconstruction algorithm more sensitive than the derivative algorithm, or vice versa.  We find empirically that independently tuning the two
parameters provides little to no advantage in the test cases we consider in
Sec. \ref{sec:numerical_tests}, so for the remainder of this work we choose the
same value of $\epsilon_{W}$ for both the WENO reconstruction and the CWENO
derivative algorithms. 

It is important to note that since the CWENO scheme computes the spatial
derivative terms using an adaptive finite difference stencil, the irrotational
nature of the gradient of these terms ($\partial_i \bm{p}_0$, where $i$ is a
spatial index) is not exactly preserved \cite{dumbser2020numerical}.  Explicitly, consider the trivial
constraint corresponding to $\partial_i \xi$; asserting that the curl of this
gradient vanishes (and specializing to the type of problems considered here, in
Minkowski spacetime with variation only in $t, x, y$) leads one to the
constraint
\begin{equation} \label{eq:curl_constraint}
0 = \partial_x \partial_y \xi - \partial_y \partial_x \xi.
\end{equation}
It is straightforward to show that discretizations of
(\ref{eq:curl_constraint}) with fixed stencils, e.g. $\partial_x X \approx
(X_{i+1,j}-X_{i-1,j})/(2 h)$ and its analogy with $i \to j$ for $\partial_y X$,
satisfy (\ref{eq:curl_constraint}) exactly.  For the CWENO scheme, however,
constraints like (\ref{eq:curl_constraint}) are only satisfied up to truncation
error in the solution, here $O(h^2)$.  That said, for large values of the
WENO/CWENO parameter $\epsilon_{W}$ the derivatives approach those coming from
a fixed stencil, and violations of (\ref{eq:curl_constraint}) vanish; for a
thorough exploration of curl-type constraint violation for the BDNK scheme, see
Sec. \ref{sec:constraint_tests}.

After the primitive variable recovery step of Sec. \ref{sec:pvr}, we compute
the spatial derivative terms $\partial_i \bm{p}_0$ across the entire grid using
(\ref{eq:CWENO_derivative}) and save them.  We then treat them in the same way
as the non-derivative terms $\bm{p}_0$, reconstructing their values at the cell
interfaces using WENO (\ref{eq:WENO_right_reconst},\ref{eq:WENO_left_reconst})
before feeding them into the numerical flux function.

\subsection{Numerical flux} \label{sec:numerical_flux}

As was mentioned in Sec. \ref{sec:fvm}, the choice of numerical flux function
is critical to the stability of shockwave solutions in a HRSC finite volume
scheme.  The wide variety of these functions fall roughly into two categories:
\textit{upwind methods} and \textit{central methods}.  Upwind schemes treat the
interface between two cells as a Riemann problem, which is solved by feeding
information about the characteristics of the PDEs into a Riemann solver.  This
procedure allows such schemes to bias the required stencils
such that they are upwind with respect to the flow, dramatically improving
stability. Central schemes, on the other hand, eschew use of detailed
characteristic information and Riemann solvers in favor of simple
discretizations with stencils centered about cell interfaces. For a
detailed assessment and discussion of central schemes in the context of
astrophysical applications see Ref. \cite{Lucas-Serrano:2004fqa}.

Both upwind and central schemes have been successfully applied to the
relativistic Euler equations.  For the BDNK equations, however, we find that
computation of the characteristic information required for an upwind
scheme---for example, computation of the linearized flux Jacobian
$\frac{\partial \bm{f}}{\partial \bm{q}}$ required in a Roe scheme
\cite{Roe_1997}---is difficult and yields a numerical flux which is
computationally expensive to evaluate.  Hence we opt for a Riemann-solver-free
central scheme, specifically one based on the Kurganov-Tadmor numerical flux
function \cite{kurganov2000new}
(using as an example the flux through the cell interface at $(x_{i+1/2}, y_{j})$):
\begin{multline} \label{eq:KT_flux}
\bm{F}_{i+1/2, j} = \frac{1}{2} \Big( \bm{f}(\bm{p}^{-}_{i+1/2, j}) + \bm{f}(\bm{p}^{+}_{i+1/2, j}) \\
- a \big[ \bm{q}(\bm{p}^{+}_{i+1/2, j}) - \bm{q}(\bm{p}^{-}_{i+1/2, j}) \big] \Big).
\end{multline}
The Kurganov-Tadmor flux requires only the primitive variables $\bm{p}_1$
computed at the cell interfaces (via WENO and CWENO, Sec.
\ref{sec:reconstruction}), the flux functions $\bm{f}$, and a the scalar
quantity $a$, defined to be the maximum wave propagation speed.  The value of
$a$ controls the amount of numerical diffusion applied at discontinuities, and
may be found empirically by changing $a \in [0, 1]$ until one strikes an
acceptable balance between sharp shock resolution ($a \to 0$) and stability ($a
\to 1$).  For BDNK theory, though, since we have chosen a frame where the
maximum propagation speed is equal to the speed of light, we know $a$ exactly
and set\footnote{The Kurganov-Tadmor flux with $a = 1$ is equivalent to the
so-called HLL flux \cite{Harten1983} as well as the local Lax-Friedrichs flux
\cite{Lax1954} when their respective maximum propagation speeds are set to unity.}
$a = 1$.  Since precise characteristic information is incorporated into the
numerical flux calculation, the method applies aspects of both central and
upwind schemes, and is sometimes referred to as a central-upwind scheme 
\cite{Rezzolla2013}.

It is important to note that the numerical flux (in our case
(\ref{eq:KT_flux})) is constructed such that it is symmetric in $\bm{p}^{-}$
and $\bm{p}^{+}$; this fact implies that the flux computed at the left side of
the interface is equal to that computed on the right side.  Physically, this
implies that all of the flux of $\bm{q}$ out of a cell must flow into
neighboring cells, and vise versa, such that the total quantity of $\bm{q}$
cannot change\footnote{In a computer simulation, the use of finite precision
floating point arithmetic results in round-off errors of order $10^{-16}$ at
double precision; these errors are typically many orders of magnitude smaller
than those due to truncation error in the solution, however.} in the absence of
sources or boundaries \cite{LeVeque_2006}.  Integrating (\ref{eq:conserv_law})
over a such a domain $\mathcal{D}$ (assumed to be 2D), one finds   
\begin{equation} \label{eq:discrete_cons}
\frac{\partial}{\partial t} \int_{\mathcal{D}} \bm{q} \, dx dy = 0,
\end{equation}
implying that the total quantity of $\bm{q} = (T^{tt}, T^{tx}, T^{ty})^{T}$ in
$\mathcal{D}$ is constant in time.  The fact that finite volume schemes preserve
(\ref{eq:discrete_cons}) exactly is known as \textit{discrete
conservation}, and is crucial to the success of such schemes in
countless applications.  In Sec. \ref{sec:constraint_tests} we check the
conservation of $\bm{q}$ across the simulation domain, and confirm that our
scheme possesses the discrete conservation property.

We have constructed our BDNK scheme such that it reduces to a HRSC finite
volume perfect fluid solver in the inviscid limit $\eta_0 \to 0$.  For the sake
of sharp comparisons between the viscous and inviscid cases in the tests that
follow, we also use $a = 1$ for the relativistic Euler equations, even though
their characteristic speeds are equal to the sound speed $|c_s| =
\sqrt{\frac{\partial P}{\partial \epsilon}} = \frac{1}{\sqrt{3}}$.  As
described above, choosing a larger value of $a$ results in slightly more
numerical viscosity in the solution; this numerical viscosity converges away
with resolution, and is always orders of magnitude smaller than the physical
viscosities shown in Sec. \ref{sec:numerical_tests}.

\subsection{Evolving $J^{a}$} \label{sec:evolving_J}
Since we have assumed a fluid with an underlying conformal symmetry, the stress-energy tensor $T^{ab}$ has no dependence on $n$, and (\ref{eq:T_cons_law}) decouples from (\ref{eq:J_cons_law}).  This implies one has greater freedom in choosing a method to solve (\ref{eq:J_cons_law}), since it cannot destabilize the solution to (\ref{eq:T_cons_law}). In fact, if one is not interested in the evolution of $n$, one may forego solving (\ref{eq:J_cons_law}) entirely and just solve (\ref{eq:T_cons_law}) to evolve $\epsilon, u^{a}$.  For non-conformal fluids, $T^{ab}$ will depend on $n$, however, and (\ref{eq:T_cons_law}-\ref{eq:J_cons_law}) will have to be solved as a coupled system of PDEs.

In this work, we choose to evolve both $T^{ab}$ and $J^{a}$ using a scheme based in the finite volume method.  This entails applying the same steps described in the past three subsections---primitive variable recovery, reconstruction, and flux computation---to (\ref{eq:J_cons_law}).  Fortunately, the simple form of the particle current (\ref{eq:Ja_0_cov}) (or equivalently (\ref{eq:J_cov},\ref{eq:J_grad_correction})) simplifies this procedure significantly.
  
Primitive variable recovery is trivial for $J^{a}$, as the one conserved variable $J^{t}$, (\ref{eq:delJ_conserv_form}), is linear in the one primitive variable $n$ (which is the only possible choice of primitive variable, since $u^{a}$ is being evolved with (\ref{eq:T_cons_law})). 
The flux terms are functions only of $n, u^{a}$, and hence one may use the same reconstruction method as described above (here WENO, see Appendix \ref{sec:WENO}) to interpolate their values to cell interfaces.  
We use the same numerical flux function for (\ref{eq:T_cons_law}) and (\ref{eq:J_cons_law}), namely the Kurganov-Tadmor flux (\ref{eq:KT_flux}) with maximum local speed $a = 1$.

\subsection{Time integration}
Here, as in \cite{Pandya_2021}, we integrate the system of PDEs
(\ref{eq:finite_vol}) in time using the total-variation-diminishing second-order Runge
Kutta algorithm known as Heun's method.  Heun's method gives the conservative
variables at the unknown advanced time level, $\bm{q}^{n+1}$, by writing
(\ref{eq:finite_vol}) as ${\dot{\bm{q}} = \bm{H}(\bm{q})}$ and applying the
following procedure:
\begin{equation}
\begin{aligned}
\hat{\bm{q}}^{n+1} &= \bm{q}^{n} + \Delta t \bm{H}(\bm{q}^{n}) \\
\bm{q}^{n+1} &= \bm{q}^{n} + \frac{\Delta t}{2} \Big[ \bm{H}(\bm{q}^{n}) + \bm{H}(\hat{\bm{q}}^{n+1}) \Big].
\end{aligned}
\end{equation}
Heun's method works by producing an estimate using a forward-Euler update step,
$\hat{\bm{q}}^{n+1}$, and then uses the known level ($\bm{q}^{n}$) and the
estimate ($\hat{\bm{q}}^{n+1}$) to find the conservative variables at the
unknown time level ($\bm{q}^{n+1}$).\\

\subsection{Boundary conditions}
In the tests that follow, we are exclusively interested in the dynamics in the
interior of the simulation domain, and the boundaries have no physical
interpretation.  That said, finite computational resources dictate that
boundaries are necessary, and we designate the outermost three grid cells in
each direction as boundary cells.  For most of the simulations described below,
we define the boundary cells to be \textit{ghost cells}, whereby the state in
the cell is set to be the same as that in the nearest non-ghost cell.
Explicitly, at the boundaries at constant $y$ (at constant values of the second
index) we take
\begin{equation} \label{eq:ghost_cells}
\begin{aligned}
X_{k, j} &\coloneqq X_{3, j}, ~~~~~~~ k \in [0, 2] \\
X_{k, j} &\coloneqq X_{N-4, j}, ~~~ k \in [N-3, N-1]
\end{aligned}
\end{equation}
for the hydrodynamic variables $\bm{p}_0$,
and $A \coloneqq B$ is shorthand for
``$A$ is set equal to $B$''.  The boundaries at constant $x$ (constant first
index) are obtained from the above after switching the indices.

The use of ghost cells is common in numerical hydrodynamics, though in this case one must also determine how to handle the derivative terms, both spatial ($\partial_i \bm{p}_0$) and temporal ($\bm{p}_1$).  It is clear that the choice for $\partial_i \bm{p}_0$ must be consistent with the choice for $\bm{p}_0$ in the ghost cells (\ref{eq:ghost_cells}), but it is not so obvious how to treat $\bm{p}_1$.  For all of the problems with ghost cells boundaries considered here, we find no real difference between using (\ref{eq:ghost_cells}) for $\bm{p}_1$ or setting $\bm{p}_1$ to zero in the ghost cell region; this is largely due to the design of the problems, however, as boundary conditions are not the main focus of this study.  We will investigate boundary conditions for the BDNK equations more thoroughly in a future work.

We also consider a test with periodic boundaries, where opposite edges of the
domain are identified.  This is achieved numerically by identifying the three
boundary cells on one side of the domain with the three non-boundary cells
nearest to the other edge of the domain, for all four edges.  Explicitly, this
procedure sets all variables $X$ along the boundary at constant $y$ (second
index) via
\begin{equation}
\begin{aligned}
X_{0, j} &\coloneqq X_{N-6, j}, ~~~ X_{1, j} \coloneqq X_{N-5, j}, ~~~ X_{2, j} \coloneqq X_{N-4, j} \\
X_{N-1, j} &\coloneqq X_{5, j}, ~~~ X_{N-2, j} \coloneqq X_{4, j}, ~~~ X_{N-3, j} \coloneqq X_{3, j}, \\
\end{aligned}
\end{equation}
where the variables along the boundaries at constant $x$ are set in the same
way as above except with the indices switched.

\section{Numerical tests} \label{sec:numerical_tests}
In this section we present the results of a series of tests which compare the
new BDNK scheme to the HRSC perfect fluid solver obtained in the inviscid limit
$\eta_0 \to 0$, as well as to the semi-finite-difference\footnote{In \cite{Pandya_2021}, the algorithm splits the stress-energy tensor into a perfect fluid piece and a dissipative correction, each of which has its own flux term.  The former is discretized using a finite-volume approach with a Roe flux \cite{Roe_1997}, and the latter with a non-conservative second-order finite difference stencil.  Since the approach of \cite{Pandya_2021} is part finite volume and part finite difference, we refer to it as a ``semi-finite-difference'' scheme.} scheme of
\cite{Pandya_2021}.  The tests are performed on either a 1D or a 2D Cartesian
grid, with variation in $(t,x)$ or $(t,x,y)$ respectively.  We define a single
grid scale $h$ in both spatial directions, and we take the difference between
time steps to be $\Delta t = \lambda h/a$, where $a = 1$ is the maximum local characteristic speed and the Courant
factors $\lambda \in (0, 1)$ for the tests are reported in Table \ref{table:courant}.

In all of the simulations below, initial data is set by prescribing values for
the hydrodynamic variables $\epsilon, n, u^{a}$; viscous corrections are
initialized to zero, so $T_1^{tc}$ is set at $t = 0$ using $T_{0}^{tc}$.

All dimensionful quantities are given in code units\footnote{We use natural units with energies measured in $\textnormal{GeV}$, which implies velocities are dimensionless $[u^{a}] = 1$, coordinates have units of inverse energies $[x^{a}] = \textnormal{GeV}^{-1}$, and thus energy densities have unit $[\epsilon] = [T^{ab}] = \textnormal{GeV}^{4}$.},
which are the same as in \cite{Pandya_2021}.
It is important to note that the following simulations are tests designed to evaluate the performance of the algorithm, not attempts to model a known physical system. The scales chosen in these tests are arbitrary, and we choose the amount of viscosity used based on whether the dynamics are underdamped or overdamped, rather than by reference to a substance where the viscosity is known. In particular, none of the initial data we consider are particularly close to that relevant in modeling heavy-ion collisions, and hence it is not that meaningful to quantify viscosities via the entropy-normalized shear viscosity $\eta/s$ as is typically done in the nuclear physics literature.  Instead, we use the parameter $\eta_0$ defined in (\ref{eq:frame_coeffs_def}).

We order the set of tests into three categories: (1) tests of constraint
preservation; (2) tests with sharp gradients; and
(3) tests of the BDNK solutions approaching the inviscid limit.  In each
section we include results from both 1D and 2D simulations.

\begin{table}[h]
\begin{tabular}{c || c | c}
\textit{Initial data} & Max. $\lambda$ & $\lambda$ used \\[0.5mm]
\hline \hline
1D Gaussian	     				& 0.5 & 0.1 \\
2D viscous rotor 				& 0.5 & 0.1 \\
1D shock tube	 				& 0.5 & 0.1 \\
2D oblique shockwave			& 0.1 & 0.1 \\
1D steady-state shockwave 		& 0.5 & 0.1 \\
2D Kelvin-Helmholtz instability & 0.5 & 0.5
\end{tabular}
\caption{Maximum stable Courant factor 
$\lambda \equiv \frac{a \Delta t}{h}$ 
(where the local characteristic speed $a = 1$; $\lambda = 0.5$ is the maximum value satisfying the CFL
condition for a 2D Cartesian grid \cite{titarev2004finite}) 
and $\lambda$ used
to make figures for each of the sets of initial data considered here.  Whenever
the scheme of \cite{Pandya_2021} is used for comparison, we take $\lambda =
0.1$ to improve stability of that scheme.  Lower than maximum Courant numbers
are used for the 2D viscous rotor test to minimize spurious reflections from
the boundary. See Sec. \ref{sec:sharp_gradients} for a discussion of the
stability of the 2D oblique shockwave test.} \label{table:courant}
\end{table}

\subsection{Tests of constraint preservation} \label{sec:constraint_tests}

\subsubsection{1D Gaussian test}

\begin{figure}[h]
	\centering
	\includegraphics[width=\columnwidth]{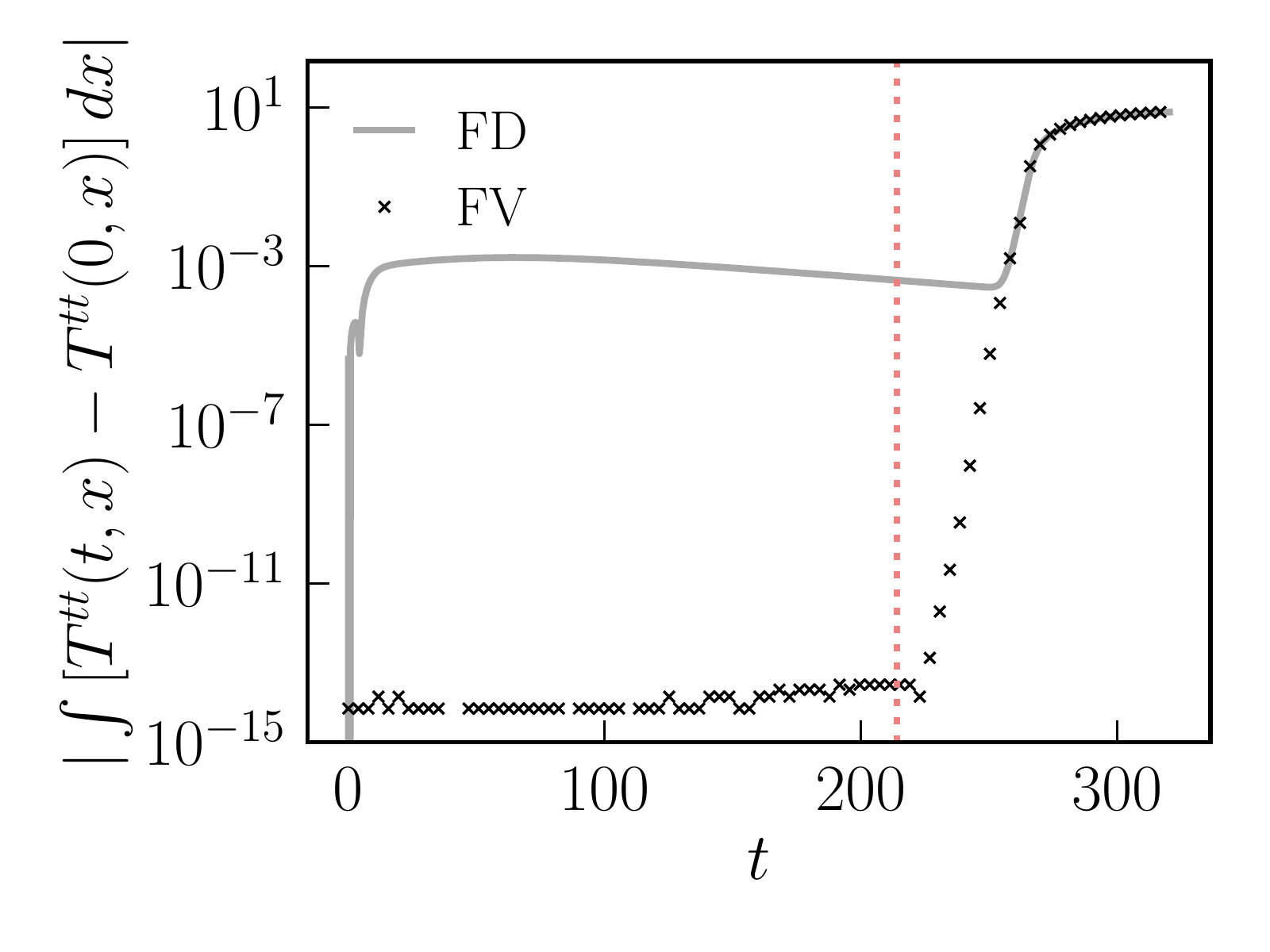}
    \caption{Discrete conservation of $T^{tt}$ across the spatial domain for a
    simulation starting from Gaussian initial data (\ref{eq:gaussian_ID}) 
    with $\eta_0 = 0.2$ for
    the finite volume scheme presented here (``FV'') as well as the semi-finite-difference scheme of \cite{Pandya_2021} (``FD'').  As expected, the finite
    volume scheme conserves $T^{tt}$ up to machine precision, $\sim 10^{-15}$,
    until the fluid pulse reaches the boundary at the time marked by the light
    red dotted line.  The semi-finite-difference scheme of \cite{Pandya_2021}
    conserves $T^{tt}$ only up to the level of truncation error, which in this
    case is $\sim 10^{-3}$.} \label{fig:Tab_cons}
\end{figure}

We will first check the ability of the new scheme to preserve the spatial
integral of the conservation law over the simulation domain in the absence of
sources or significant boundary interactions (\ref{eq:discrete_cons}).  To do
so, we first consider the simplest possible test, namely a 1D simulation
starting from smooth initial data in $x$, as in \cite{Pandya_2021}.
Explicitly, at the initial time we take a stationary Gaussian profile in the
energy density
\begin{equation} \label{eq:gaussian_ID}
\epsilon(t=0, x) = A e^{-x^2/w^2} + \delta, ~~~ u^{x}(t=0, x) = 0,
\end{equation}
with parameter values $A = 1, w = 25, \delta = 10^{-1}$, and we take the
simulation domain to be $x \in [-L, L]$, where $L = 200$.  For the viscosity
we choose $\eta_0 = 0.2$.
Since it is smooth, the initial
data (\ref{eq:gaussian_ID}) gives results which are very similar to those given
in \cite{Pandya_2021}.  The key difference, however, is that since the new
scheme is conservative, the integrals of motion
(\ref{eq:discrete_cons}) are conserved to machine precision at times when no fluid is leaving the
boundaries of the domain 
(\ref{eq:discrete_cons}); the semi-finite-difference scheme
of \cite{Pandya_2021} conserves them only to truncation error, which is roughly
12 orders of magnitude larger---see Fig. \ref{fig:Tab_cons}.

\subsubsection{2D viscous rotor} \label{sec:viscous_rotor}

\begin{figure*}
	\centering
	\includegraphics[width=0.7\textwidth]{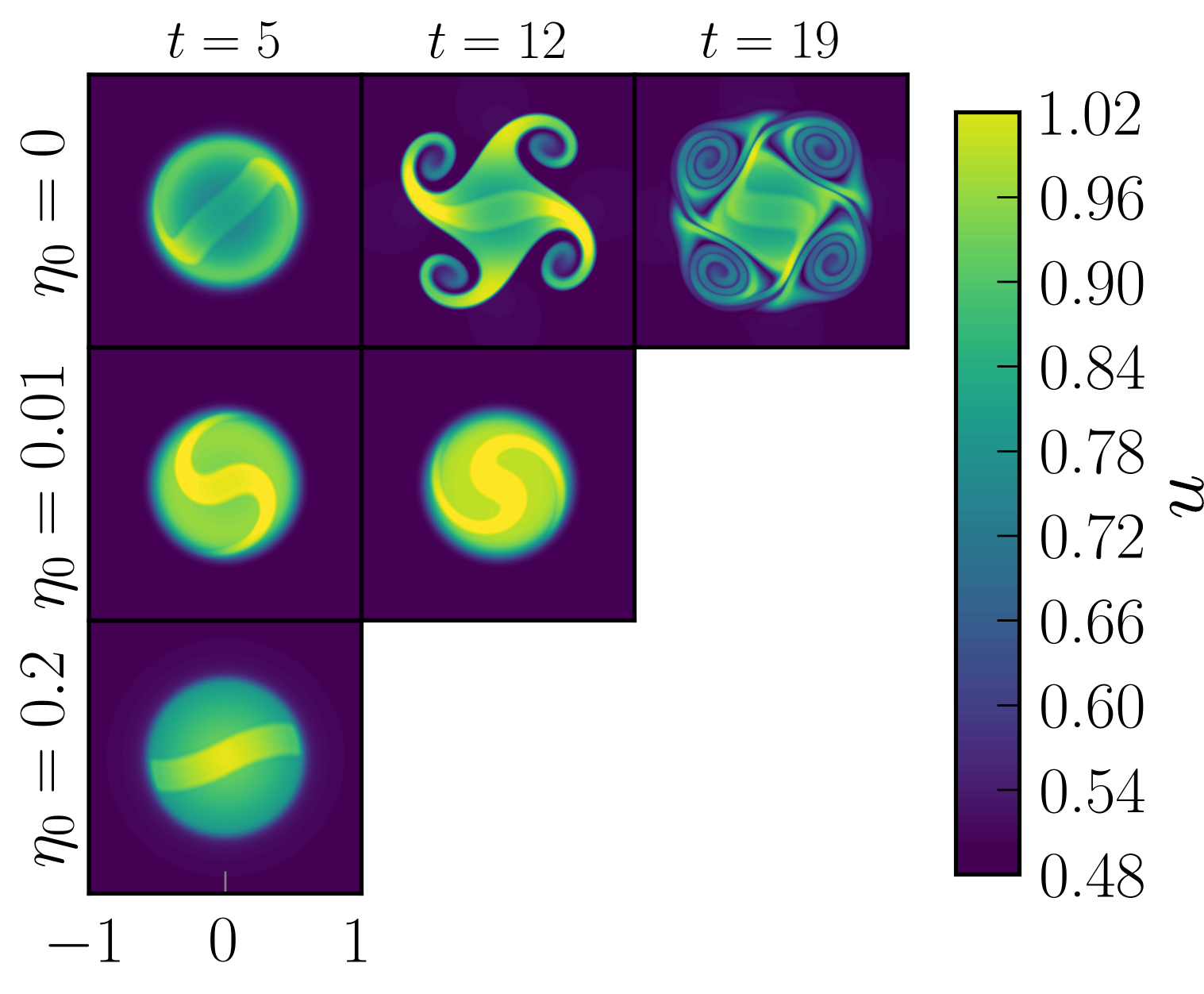}
    \caption{Density ($n$) evolution of viscous rotor initial data
    (\ref{eq:viscous_rotor_ID}) as a function of time (columns) for three
    different viscosities: $\eta_0 = 0, 0.01, 0.2$ in rows, from top to bottom.
    In the inviscid simulation, the cylinder of fluid is Kelvin-Helmholtz
    unstable and forms vortices which are not present in the viscous cases.  At
    intermediate viscosity (middle row), the fluid experiences a shearing force
    which distorts the bar of overdensity present in the initial data, before
    the cylinder stops rotating entirely around $t \sim 12$.  At the highest
    viscosity shown (bottom panel), the cylinder rotates only about
    $20^{\circ}$ before stopping at $t \sim 5$. 
} \label{fig:rotor}
\end{figure*}
In tests with two or more spatial dimensions, one must be careful to preserve
``curl''-type constraints, of which (\ref{eq:curl_constraint}) is an example.
These constraints are satisfied exactly when derivative terms are approximated
using fixed stencils; such stencils are unstable about sharp gradients,
however, so we opt for an adaptive scheme based upon the CWENO algorithm (see
Sec. \ref{sec:reconstruction}).

\begin{figure}
	\centering
	\includegraphics[width=\columnwidth]{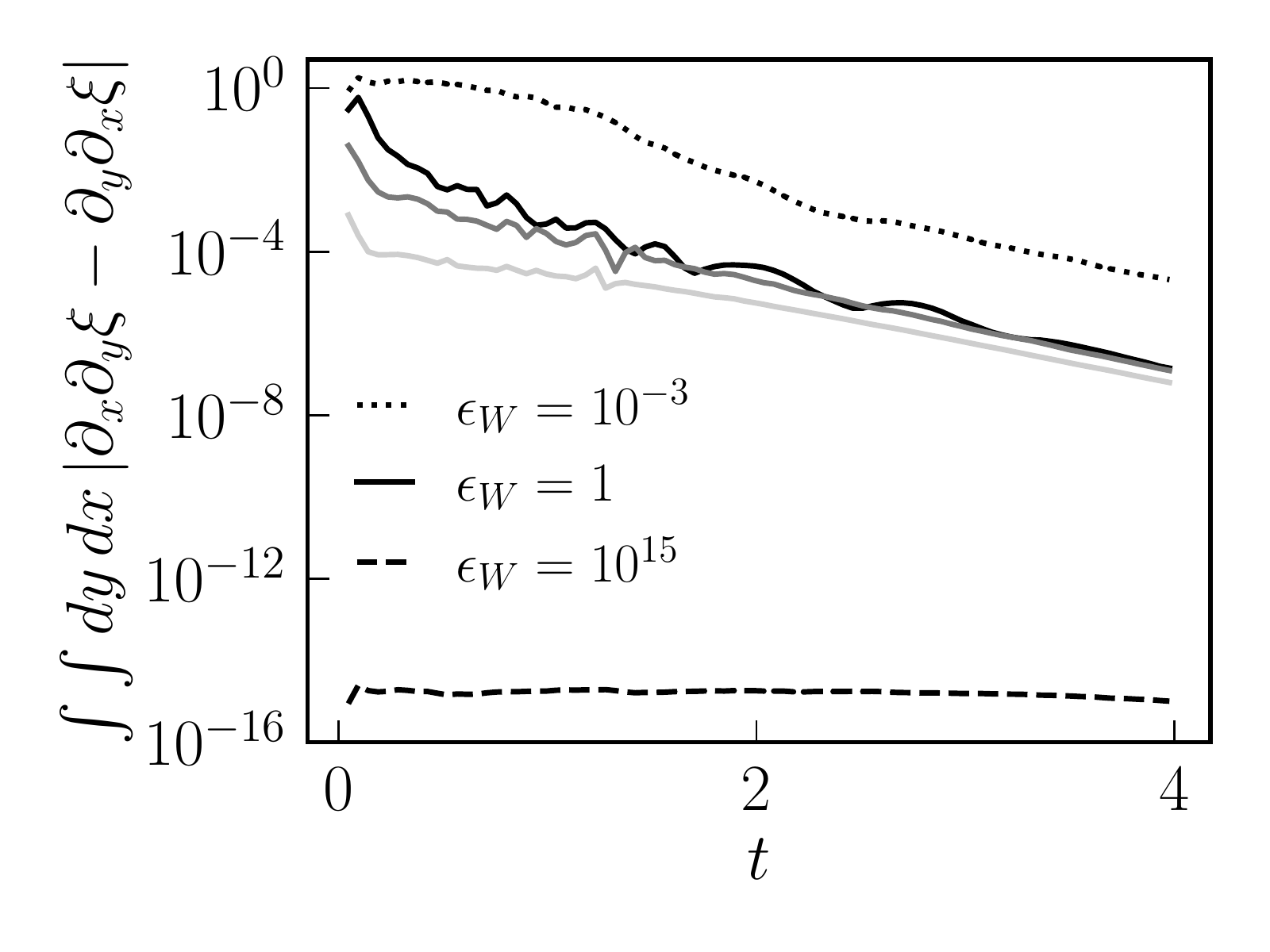}
    \caption{Integral of the absolute value of the constraint
    (\ref{eq:curl_constraint}) over the domain for viscous rotor initial data
    with $\eta_0 = 0.2$.  The value of $\epsilon_{W}$ determines how strongly the
    smoothness of the candidate ENO stencils impacts the nonlinear weights;
    small values of $\epsilon_{W}$ imply strong sensitivity to nonsmoothness,
    and large values imply insensitivity (and as a result give a fixed
    fourth-order derivative stencil).  Hence, for smaller $\epsilon_{W}$ one
    finds larger violations of the constraint (\ref{eq:curl_constraint}), which
    converge away with resolution (the solid lines range over $N_x = 2^7, 2^{8}, 2^{9}$, with lighter shades representing higher resolutions).  In the $\epsilon_{W} \to
    \infty$ limit, constraint violation approaches machine precision (cf. the
    $\epsilon_{W} = 10^{15}$ case).} \label{fig:CC_plot}
\end{figure}

To evaluate the effectiveness of the new scheme at preserving constraints like
(\ref{eq:curl_constraint}), we study a set of initial data which corresponds to
a ``viscous rotor'', namely a fluid at constant pressure where a cylinder in
the center of the domain is initially rotating at constant angular velocity
$\omega$.  We implement the viscous rotor initial data on a grid with $x, y \in
[-L, L]$ with $L = 3$, where at $t = 0$
\begin{equation} \label{eq:viscous_rotor_ID}
\begin{aligned}
\epsilon(x,y) &= 1 \\
v^{x}(x, y) &= -\omega \sqrt{x^2 + y^2} \sin(\theta) D(d, \delta) \\
v^{y}(x, y) &= \omega \sqrt{x^2 + y^2} \cos(\theta) D(d, \delta) \\
n(x, y) &= \frac{1}{2} \Big( D(d, \delta) + 1 \Big) + B(x, y) \\
\end{aligned}
\end{equation}
which gives a fluid at constant pressure $P = \frac{1}{3}$, with a circular
region in the center initially rigidly rotating with angular velocity $\omega =
1$.  This is implemented via functions
\begin{equation}
\begin{aligned}
\theta &\equiv \textnormal{atan2}(y,x) \\
D(d, \delta) &\equiv \frac{1}{2} \Big[1 + \tanh \Big(\frac{d}{\delta} \Big) \Big] \\
d &\equiv R - \sqrt{x^2 + y^2} \\
B(x, y) &= 
\begin{cases}
0.1 & \sqrt{x^2 + y^2} \leq R ~ \& ~ |y| < 0.1 \\
0 & \textnormal{otherwise}
\end{cases}
\end{aligned}
\end{equation}
where $\textnormal{atan2}(y,x)$ is the two-argument arctangent, $D(d, \delta)$
is a function which is unity at the origin and decreases sharply but smoothly
at radius $R = 0.5$, with the smoothness of the transition controlled by
$\delta$, which we take to be $0.05$. $B(x, y)$ gives a raised bar oriented horizontally in the
center of the rotating region which may be used to see how far the rotor has
spun.  

Fig. \ref{fig:rotor} shows the evolution of this set of initial data as a
function of time (columns) and viscosity (rows).  From the figure, it is
immediately clear that the viscosity has a significant effect on the late-time
behavior of the solution---the inviscid case continues rigidly rotating for a
while, leaving the bar of overdensity approximately straight up until
the solution becomes Kelvin-Helmholtz unstable and forms
vortices\footnote{Circular symmetry is broken by the square grid, and the
grid-scale bumps at the top, bottom, leftmost, and rightmost points on the
circle each source the Kelvin-Helmholtz instability.  Convergence is typically
lost after these vortices form, as both the size of the perturbation and the
numerical viscosity in the solution decrease as the grid is refined.}; the
intermediate viscosity case experiences a strong shearing force, distorting the
bar into an ``S''-shape before stopping at $t \sim 12$; and the high viscosity
case stops almost immediately after $t \sim 5$.

Fig. \ref{fig:CC_plot} shows violations of the constraint
(\ref{eq:curl_constraint}) for the case with $\eta_0 = 0.2$, where the viscous
contribution to the fluxes is significant.  Plotted in the figure are a set of
lines with varying values of the WENO/CWENO parameter $\epsilon_{W}$, which
determines the amount of ``stencil switching'' that occurs during a simulation.
As is described in Sec. \ref{sec:reconstruction} and App. \ref{sec:WENO}, at low values of
$\epsilon_{W}$ the CWENO algorithm adjusts the nonlinear weights to be
significantly different from the linear ones, producing a non-uniform stencil
and consequently violating the constraint (\ref{eq:curl_constraint}); these
violations converge away with numerical resolution, however (shown in the
solid lines of varying shade for $\epsilon_{W} = 1$).  In the limit
$\epsilon_{W} \to \infty$ (approximated in the figure with $\epsilon_{W} =
10^{15}$), the CWENO algorithm gives a fixed, fourth-order centered finite
difference stencil, and the violation of (\ref{eq:curl_constraint}) drops to
near machine precision.

\subsection{Tests with sharp gradients} \label{sec:sharp_gradients}

\subsubsection{1D shock tube}
Though it remains unclear whether discontinuous solutions are sensible in BDNK
theory\footnote{It is well known that the weak formulation of the relativistic Euler equations possesses discontinuous solutions which are typically used to model shockwaves.  In BDNK theory, there is evidence that one should expect shockwave solutions to be smooth for ``good'' hydrodynamic frames \cite{Pandya_2021,freistuhler2021nonexistence}, potentially eliminating the physical need for discontinuous solutions.  Furthermore, the presence of derivative terms in the BDNK stress-energy tensor complicates the formulation of the Riemann problem, which has yet to be solved for BDNK theory.},
one is still free to pose discontinuous initial data; such states may
be interpreted as smooth solutions that are unresolved at the current grid
resolution.  It is essential that our algorithm be able to capture solutions
with unresolved shockwaves, as such features are prevalent in astrophysics
applications.

\begin{figure*}
	\centering
	\includegraphics[width=0.8\textwidth]{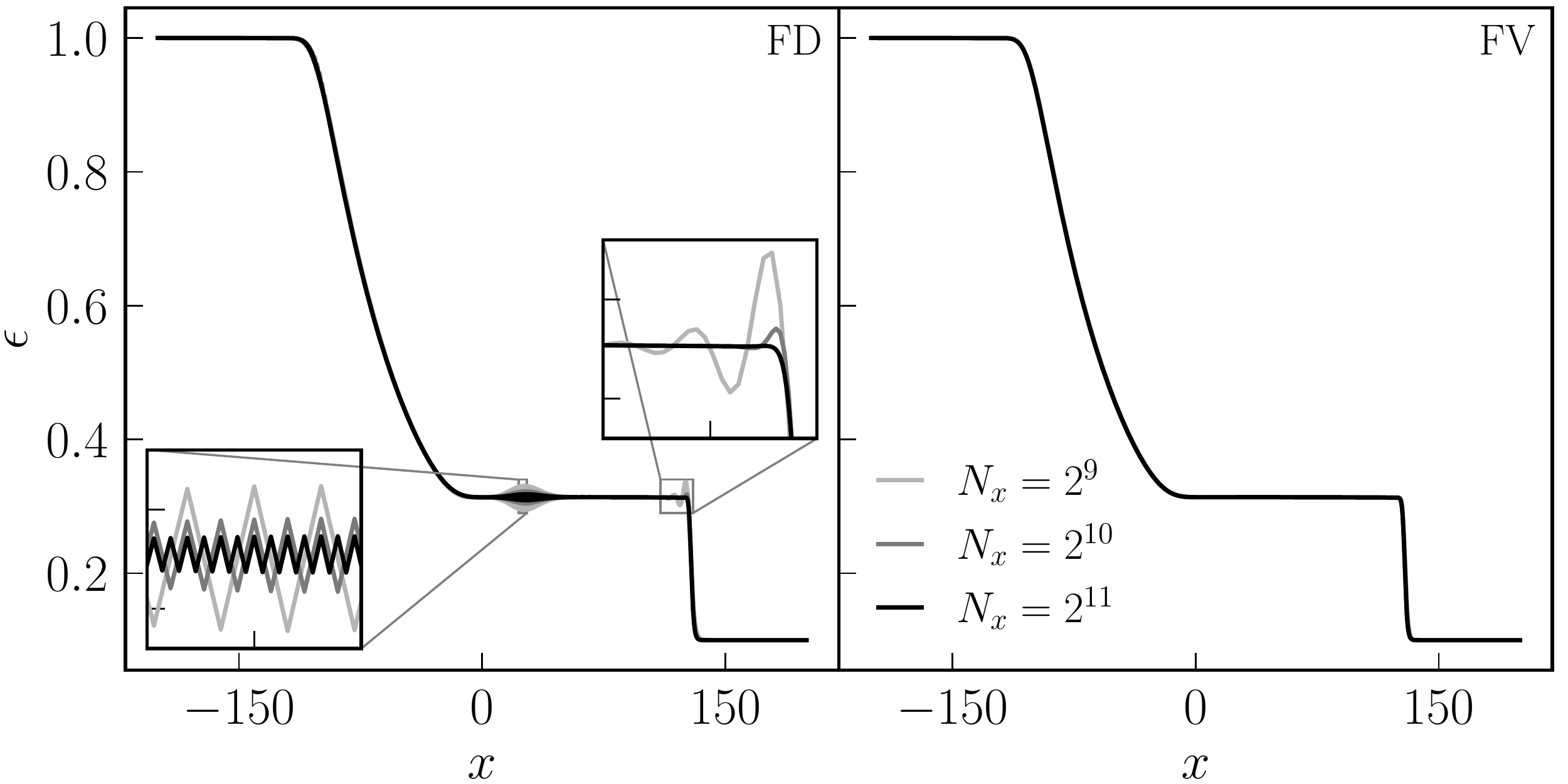}
    \caption{Comparison of solutions for $\epsilon$ starting from shock tube
    initial data (\ref{eq:1D_shock_tube_ID}) at three successive resolutions
    for the semi-finite-difference scheme of \cite{Pandya_2021} (left panel, ``FD'') versus
    the finite volume scheme presented here (right panel, ``FV'') at $t \sim 43$ for
    $\eta_0 = 0.2$.  The ``FD'' scheme has oscillations near the shock front which
    quickly converge away with resolution, as well as grid-scale ``sawtooth''
    oscillations that developed early on near the origin (the $t=0$ location
    of the shock front) and do not converge away as rapidly with resolution.
    These features do not appear in the figures of \cite{Pandya_2021} because the discontinuities there are smaller in amplitude, leading to oscillations small enough to be tamed by applying Kreiss-Oliger dissipation; said dissipation is not strong enough to remove the oscillations for the initial data (\ref{eq:1D_shock_tube_ID}), and we choose not to apply artificial dissipation in either scheme throughout this work.
    The ``FV'' solutions are free of noticeable oscillations, and the $N_x = 2^{9}, 2^{10}, 2^{11}$ curves all overlap at the resolution of the plot.}
    \label{fig:1D_shock}
\end{figure*}

To evaluate the performance of our algorithm for solutions with sharp
gradients, we first consider the standard 1D shock tube test on a domain
with $x \in [-L, L]$ with $L = 200$, where
\begin{equation} \label{eq:1D_shock_tube_ID}
\epsilon(t=0, x) =
\begin{cases}
1 & x \leq 0\\
0.1 & x > 0,
\end{cases}
~~~~~
u^{x} = 0
\end{equation}
and $\eta_0 = 0.2$, again following \cite{Pandya_2021} except with a larger difference between the
left and right states.  This set of initial data highlights the advantages of a
non-oscillatory, conservative discretization over the semi-finite-difference
discretization of \cite{Pandya_2021} in that the former gives a stable,
convergent evolution, and the latter is plagued by spurious oscillations which
do not quickly converge away (see Fig. \ref{fig:1D_shock}).

\subsubsection{2D oblique shockwave}

Outside of one spatial dimension, it is now possible for the fluid 
to possess sharp gradients which are not aligned with the numerical grid.  To
test this scenario we adopt the 2D oblique shockwave initial data of
\cite{East_2012}, whereby the simulation domain is divided into four regions:
\begin{equation}
(n, P, v^x, v^y) = 
\begin{cases}
(0.5, 1, 0, 0) & x < 0, y < 0 \\
(0.1, 1, 0, 0.97) & x > 0, y < 0 \\
(0.1, 1, 0.97, 0) & x < 0, y > 0 \\
(0.1, 0.01, 0, 0) & x > 0, y > 0.
\end{cases}
\end{equation}
Since the oblique shockwave forms dynamically during the simulation, we find it
unnecessary to use discontinuous initial data, which may be ill-posed for BDNK
theory.  Hence we follow the pattern of tests described earlier and adopt a
smoothed version of this set of initial data with tunable sharpness parameters.
We use a grid with $x, y \in [-L, L]$ with $L = 200$, and define the initial
data by
\begin{equation} \label{eq:2D_shock_ID}
\begin{aligned}
n &=0.4 \, D(d_{n}, \delta) + 0.1 \\
\epsilon &= 3 - 2.97 \, D(d_{\epsilon}, \delta) \\
v^{x} &= 0.97 \, D(d_{v^{x}}, \delta) \\
v^{y} &= 0.97 \, D(d_{v^{y}}, \delta),
\end{aligned}
\end{equation}
with
\begin{equation}
\begin{aligned}
d_{n} &= L - \Big[ (x + L)^{\gamma} + (y + L)^{\gamma} \Big]^{\frac{1}{\gamma}} \\
d_{\epsilon} &= L - \Big[ (x - L)^{\gamma} + (y - L)^{\gamma} \Big]^{\frac{1}{\gamma}} \\
d_{v^{x}} &= L - \Big[ (x + L)^{\gamma} + (y - L)^{\gamma} \Big]^{\frac{1}{\gamma}} \\
d_{v^{y}} &= L - \Big[ (x - L)^{\gamma} + (y + L)^{\gamma} \Big]^{\frac{1}{\gamma}}, \\
\end{aligned}
\end{equation}
where $\delta$ controls the smoothness of the transitions and $\gamma$ controls the squareness of each quadrant; we set $\gamma = \delta = 10$ here.  
This set of initial data is
designed such that shortly after $t = 0$, high-velocity flows from the upper
left and lower right quadrants meet the high pressure flow from the lower left
quadrant; the result is a high pressure, high velocity flow, bounded by an
almond-shaped shockwave, which propagates through the low-pressure upper right
quadrant.  

\begin{figure}
	\centering
	\includegraphics[width=\columnwidth]{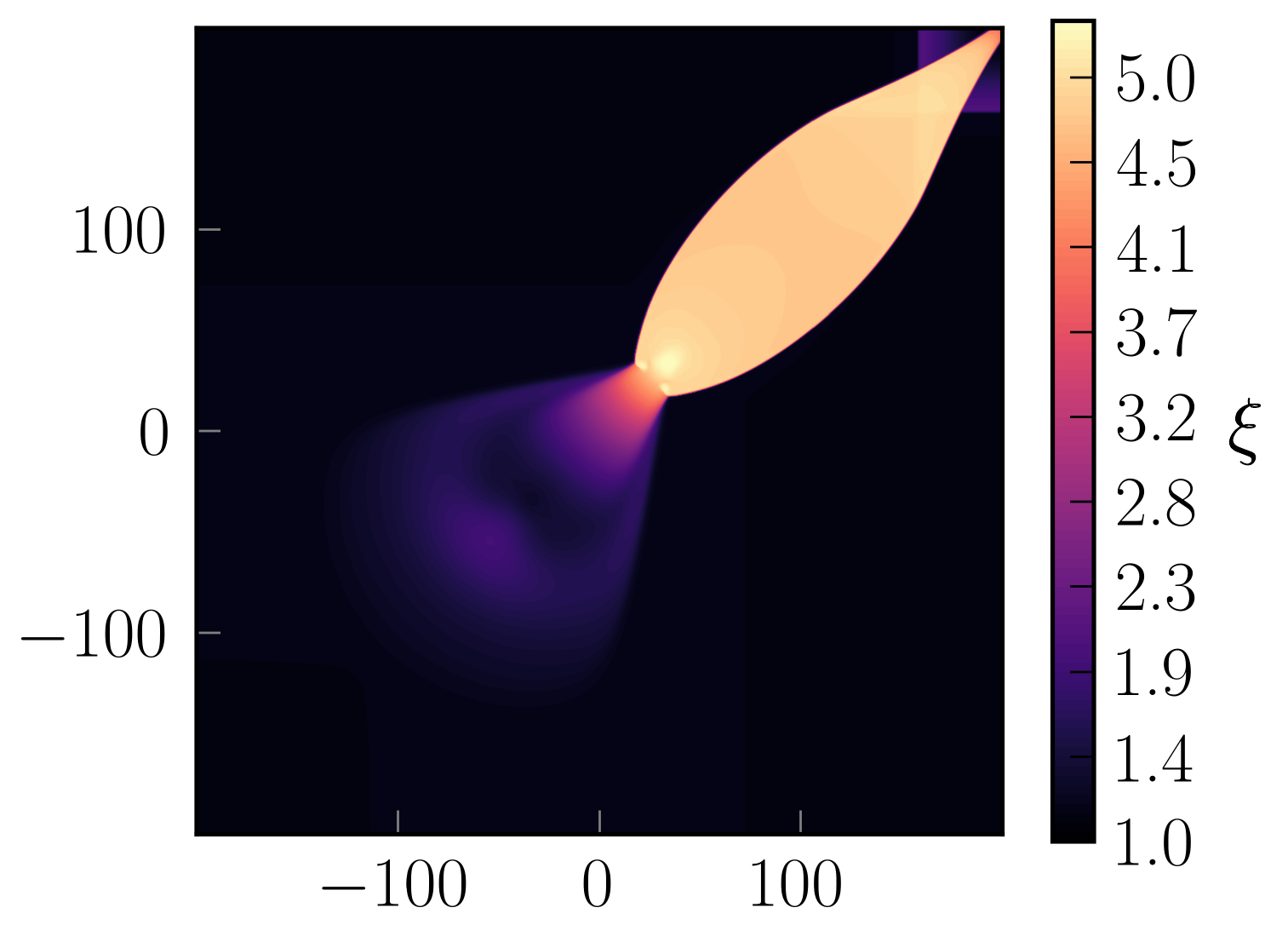}
    \caption{Solution for the log of the energy density, $\xi$, for the 2D
    oblique shockwave initial data (\ref{eq:2D_shock_ID}) at $t \sim 220$ for
    $\eta_0 = 0.2$.  Note that the solution is non-oscillatory, even though there
    is an order unity jump in $\xi$ (corresponding to a jump
    of $\sim 50$ in $\epsilon$) which is not aligned with the numerical
    grid. 
}
    \label{fig:2D_step}
\end{figure}

The non-grid-aligned shockwave can be a significant source of spurious
numerical oscillations; fortunately, these can be managed by careful choice of
the WENO/CWENO parameter $\epsilon_{W}$.  For the case shown in Fig.
\ref{fig:2D_step}, which has viscosity $\eta_0 = 0.2$, we use $\epsilon_{W} = 1$; significantly larger values of
$\epsilon_{W}$ do not choose stencils which avoid the shock, and are prone to
oscillations, and significantly smaller values of $\epsilon_{W}$ switch
stencils too frequently, introducing oscillations into the derivative terms
found in the viscous fluxes.  That said, for the case shown in Fig.
\ref{fig:2D_step} the solution is largely free of oscillations.

This set of initial data (\ref{eq:2D_shock_ID}) is posed as a challenging code
test and as such it clearly illustrates the limitations of our current
algorithm, which crashes for $v^{x}, v^{y} \gtrsim 0.98$.  In these cases, the
solution is stable until a while after the shockwave forms; the instability
sets in near the ``base of the almond'', around the origin, where numerical
error leads (\ref{eq:shifted_BDNK_pvr}) to produce a complex result, crashing
the code.  Stability can be restored for higher initial velocities $v^{x},
v^{y} \sim 0.98$ by significantly reducing the Courant factor to $\lambda =
0.05$ or even $0.01$; unfortunately, these values would likely be prohibitively
expensive at higher resolutions or in 3D simulations.  That said, the fact that
the solutions are stabilized by reducing $\lambda$ implies that the dominant
error is coming from the time integration algorithm, and these simulations
may be rendered stable by use of a higher order time integration scheme in
place of the second-order one used here.

\subsection{Tests of the inviscid limit} \label{sec:results_inv_limit}

\subsubsection{1D steady-state shockwave}
For a clear illustration of the behavior of our algorithm in the inviscid
limit---which is designed to use the BDNK primitive variable solution only in
regions where the physical viscosity is resolved---we now consider a case which
has a clearly defined equilibrium region (where gradients are negligible and
the perfect fluid approximation is valid) as well as a non-equilibrium region
(where viscous corrections are appreciable).  Specifically, we consider the
case of a planar shockwave in its rest frame, as discussed in
\cite{Pandya_2021}.  This solution is characterized by a central, smooth shockwave
bridging the transition between two equilibrium states at $x \to \pm \infty$.

To model this shockwave, we choose a set of initial data which asymptotically
(as $x \to \pm \infty$) satisfies the Rankine-Hugoniot
conditions\footnote{These conditions may be straightforwardly derived from the
relativistic Euler equations by assuming a solution independent of time; see
\cite{Pandya_2021}.} for an ideal fluid shockwave in its rest frame: 
\begin{equation} \label{eq:rankine_hugoniot}
\begin{aligned}
\epsilon_{R} &= \epsilon_{L} \, \frac{9 v_{L}^2 - 1}{3(1 - v_{L}^2)} \\
v_{R} &= \frac{1}{3 v_{L}}, \\
\end{aligned}
\end{equation}
where the equilibrium state to the left of the shockwave has parameters
$\epsilon = \epsilon_{L}, v^{x} = v_{L}$, and the state on the right is defined by
$\epsilon = \epsilon_{R}, v^{x} = v_{R}$. One should expect (\ref{eq:rankine_hugoniot}) to
describe the analogous BDNK shockwave solution as well, provided one is
sufficiently far outside of the shock.

Inside the shock, the viscous terms in the BDNK equations should produce a
smooth profile transitioning between the two states.  This precise profile is
not known analytically, but we find that choosing a set of initial data which
is sufficiently close to this profile leads to a solution which quickly settles
down to the desired steady-state shockwave solution.  For this initial data we
choose a setup with left and right states given by (\ref{eq:rankine_hugoniot}),
and the following smooth transition between the two states at $x = 0$
(here given by the Gaussian error function, $\textnormal{erf}(x)$):
\begin{equation} \label{eq:steady_state_ID}
\begin{aligned}
\epsilon(t=0, x) &= \frac{\epsilon_{R}-\epsilon_{L}}{2} \Big[ \textnormal{erf}\Big(\frac{x}{w}\Big) + 1 \Big] + \epsilon_{L} \\
v^{x}(t=0, x) &= \frac{v_{R}-v_{L}}{2} \Big[ \textnormal{erf}\Big(\frac{x}{w}\Big) + 1 \Big] + v_{L} \\
\end{aligned}
\end{equation}
where $w = 10$.  We choose the left state to be given by $\epsilon_{L} = 1, v_{L} = 0.8$, and
the right state is then computed using (\ref{eq:rankine_hugoniot}).  The
evolution quickly reaches the steady-state solution after a small blob of fluid
propagates out of the domain, changing the shock profile from the
$\textnormal{erf}$ function to one that satisfies the BDNK equations in the
static limit (see \cite{Pandya_2021} Appendix C).

\begin{figure}[h]
	\centering
	\includegraphics[width=\columnwidth]{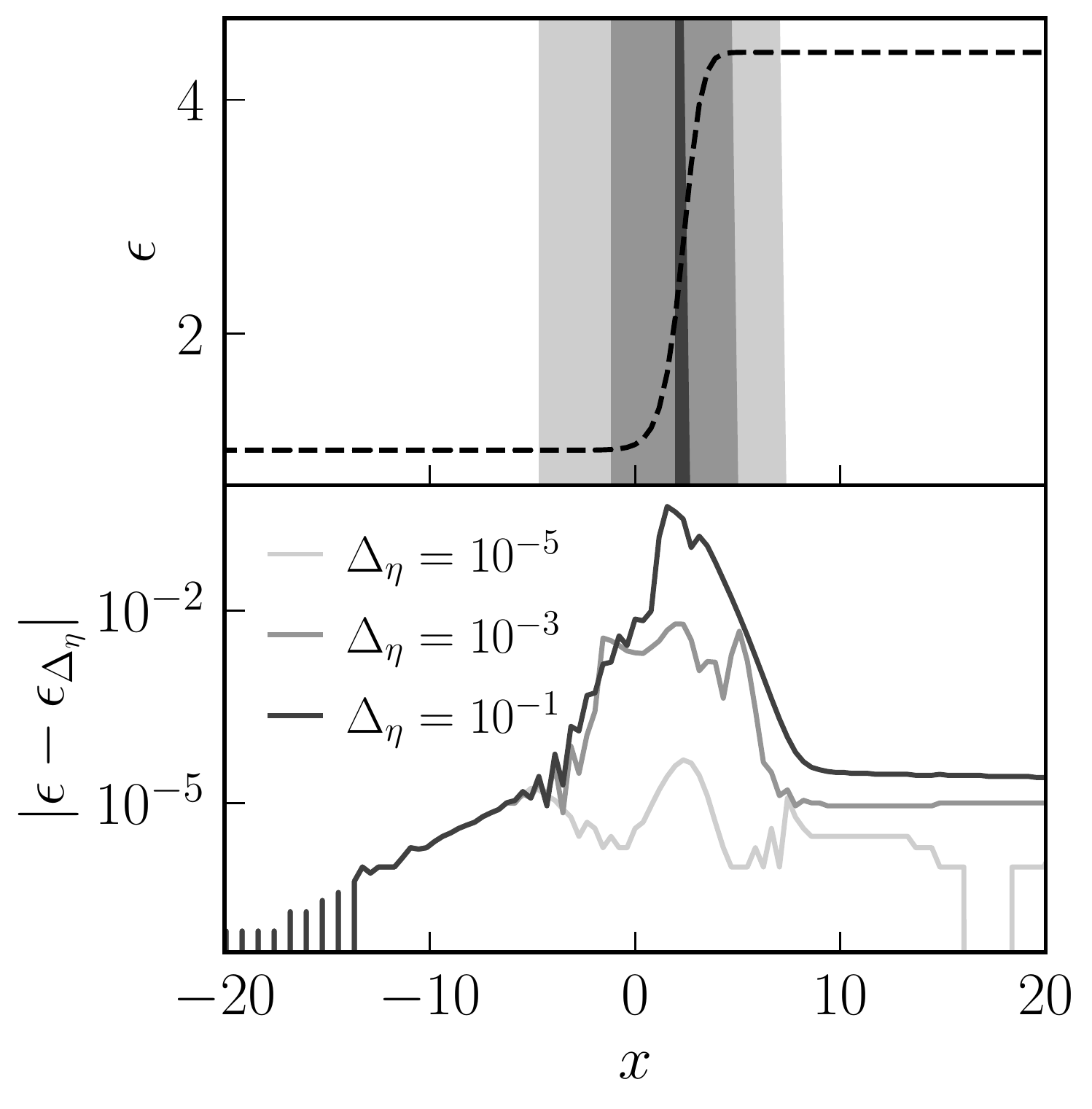}
    \caption{Illustration of the algorithm used to capture the perfect fluid
    limit for steady-state shockwave initial data (\ref{eq:steady_state_ID}) at $\eta_0 = 0.2$.
    The dashed line in the top panel is the solution for this set of initial
    data at late times, constructed using the BDNK primitive
    variable solution (\ref{eq:shifted_BDNK_pvr}) everywhere.  When the adaptive primitive variable solver is
    used, (\ref{eq:shifted_BDNK_pvr}) is only used in the gray region,
    where the shade of gray corresponds to the value of the viscous tolerance
    $\Delta_\eta$ shown in the legend.  For large values of this
    tolerance, only regions with very steep gradients are identified as being
    non-equilibrium, and (\ref{eq:shifted_BDNK_pvr}) is only used in a
    small sliver of the solution (and the perfect fluid primitive variable solution (\ref{eq:PF_pvr}) is
    used elsewhere).  This induces significant errors (bottom panel) when compared to the
    solution where only (\ref{eq:shifted_BDNK_pvr}) is used.  
    Shrinking the viscous tolerance $\Delta_\eta$ results
    in more of the non-equilibrium region being identified as such by the
    algorithm, and gives successively smaller errors when compared to the
    dashed (BDNK-only) solution.  For $\Delta_\eta \lesssim 10^{-7}$, the error
    drops to machine precision.} \label{fig:steady_state}
\end{figure}

The steady-state shock profile for $\eta_0 = 0.2$ is shown in the top panel of
Fig. \ref{fig:steady_state} as a dashed black line.  At this viscosity and
resolution, the BDNK primitive variable solution (\ref{eq:shifted_BDNK_pvr}) is
stable across the entire grid; we compare the results of the adaptive algorithm
(Sec. \ref{sec:pvr}) for various tolerances $\Delta_\eta$ against this
solution.  In the top panel, the region designated as ``non-equilibrium'' is
highlighted in gray, where the shade is determined by the viscous tolerance
$\Delta_\eta$ shown in the legend.  For large $\Delta_\eta$ (darkest gray), the
algorithm only sees regions with very steep gradients as non-equilibrium, using
the perfect fluid primitive variable solution (\ref{eq:PF_pvr}) over most of
the grid.  This results in sizeable errors (bottom panel) when compared to the
true solution, where (\ref{eq:shifted_BDNK_pvr}) is used everywhere.  Using
small $\Delta_\eta$ results in more of the shockwave being designated as
``non-equilibrium'', and the error is significantly reduced.

The behavior shown in Fig. \ref{fig:steady_state} illustrates that the adaptive
primitive variable algorithm is correctly identifying the equilibrium and
non-equilibrium regions, and demonstrates the effect of the tolerance
$\Delta_\eta$ on the solution.  That said, for the case shown one is best
served by simply using the BDNK solution (\ref{eq:shifted_BDNK_pvr}) everywhere,
since it is stable; the next section shows an example where it is unstable, and
one must use the adaptive algorithm to produce a solution at the given
viscosity and numerical resolution.

\subsubsection{2D Kelvin-Helmholtz instability}

\begin{figure*}
	\centering
	\includegraphics[width=0.9\textwidth]{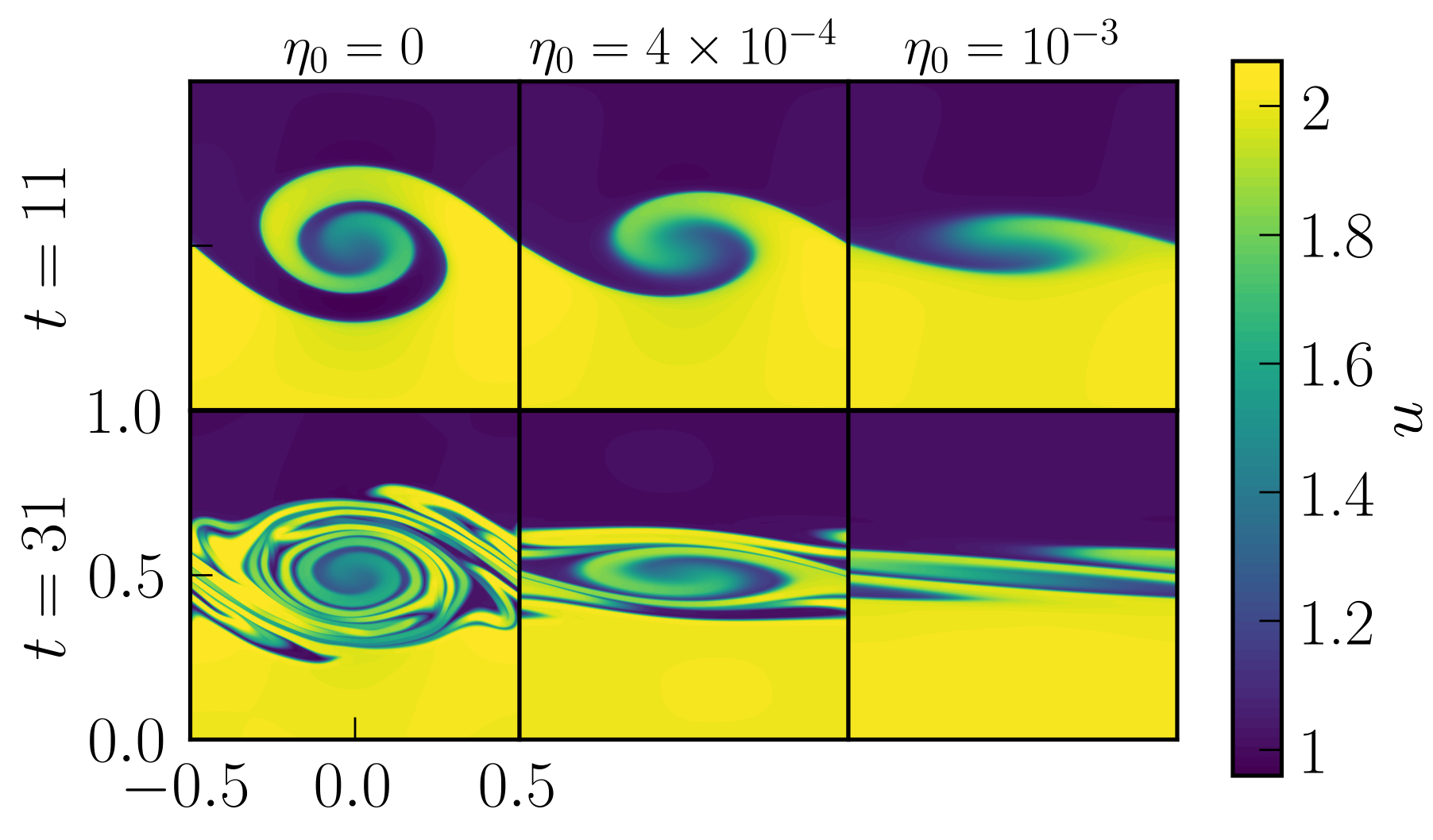}
    \caption{Evolution of Kelvin-Helmholtz-unstable initial data
    (\ref{eq:KH_ID}) for the density $n$ at three different viscosities in columns, from left to
    right: $\eta_0 = 0, 4 \times 10^{-4}, 10^{-3}$, at $t = 11$ (top row) and $t = 31$
    (bottom row).  Viscosity has a clear effect on both the early- and late-time
    state of the fluid; at $t = 11$ it determines the amount of growth of the
    perturbation of low-density fluid (dark blue) into the high-density (yellow)
    region.  For the two lower viscosity cases (left two columns), long-lived
    vortices form out of these perturbations.  At high viscosity, no clear
    vortex has formed, instead the perturbation has been sheared into a long,
    thin mixed layer. 
    } \label{fig:KH}
\end{figure*}

In this section we consider a scenario in which one may be interested in
physical viscosities which are unresolved at typical numerical resolutions,
wherein the standard BDNK primitive variable solution
(\ref{eq:shifted_BDNK_pvr}) is unstable and one requires an alternative method.
Specifically, we consider the set of initial data from \cite{Lecoanet_2015},
\begin{equation} \label{eq:KH_ID}
\begin{aligned}
\epsilon &= 1 \\
n &= 1 + \frac{1}{2}  \Big[ \tanh \Big(\frac{y-y_1}{a}\Big) - \tanh \Big(\frac{y-y_2}{a}\Big) \Big] \\
v^{x} &= u_{\textnormal{flow}} \Big[ \tanh \Big(\frac{y-y_1}{a}\Big) - \tanh \Big(\frac{y-y_2}{a}\Big) - 1 \Big] \\
v^{y} &= A \sin(2 \pi x) \Big[ \exp\Big( - \Big[\frac{y-y_1}{\sigma} \Big]^{2} \Big) \\
&~~~~~~~~~~~~~~~~~~~+ \exp\Big( - \Big[\frac{y-y_2}{\sigma} \Big]^{2} \Big) \Big],
\end{aligned}
\end{equation}
where the domain is $x \in [-L, L], y \in [-2L, 2L]$, and
$u_{\textnormal{flow}} = \frac{1}{4} c_{s} = \frac{1}{4 \sqrt{3}}, A = 0.01, a
= 0.05, \sigma = 0.2, y_{1} = -0.5, y_{2} = 0.5$.  Since the domain is twice as
large in the $y$ direction, we double the numerical resolution in that
direction, $N_y = 2 N_x$.  This set of initial data corresponds
to a jet of high density passing through a region of lower ambient density,
forming two interfaces.  These interfaces are seeded with a small perturbation
of low density into the jet region, which grows as a result of the
Kelvin-Helmholtz instability, eventually forming vortices if the viscosity of
the fluid is sufficiently small.  

Fig. \ref{fig:KH} shows snapshots from the evolution of this set of initial
data (\ref{eq:KH_ID}) for three resolutions, from left to right: $\eta_0 =  0, 4 \times 10^{-4}, 10^{-3}$, at two times (shown in columns).  Since the initial data (\ref{eq:KH_ID})
has a reflect-and-shift symmetry \cite{Lecoanet_2015} between the regions $y > 0$ and $y < 0$, only
$y \geq 0$ is shown in the figure. The effect of viscosity is
readily apparent at early times (top row), as the size to which the initial
perturbation grows (roughly, the number of winds in the spiral) diminishes with
increasing viscosity.  At late times the behavior is markedly different
between the leftmost and rightmost columns: a vortex persists for $\eta_0 = 0$,
and for $\eta_0 = 10^{-3}$ the feature from the top panel gets sheared into a
wide, thin layer.  To investigate the transition between these two disparate
behaviors, one must consider an intermediate viscosity, like that shown in the
middle column of Fig. \ref{fig:KH}.  
There, the BDNK primitive
variable solution is unstable for $N_{x} \lesssim 2^{9}$, so this case
serves as a suitable test for the adaptive primitive variable solver of Sec.
\ref{sec:pvr}.

\begin{figure}
	\centering
	\includegraphics[width=\columnwidth]{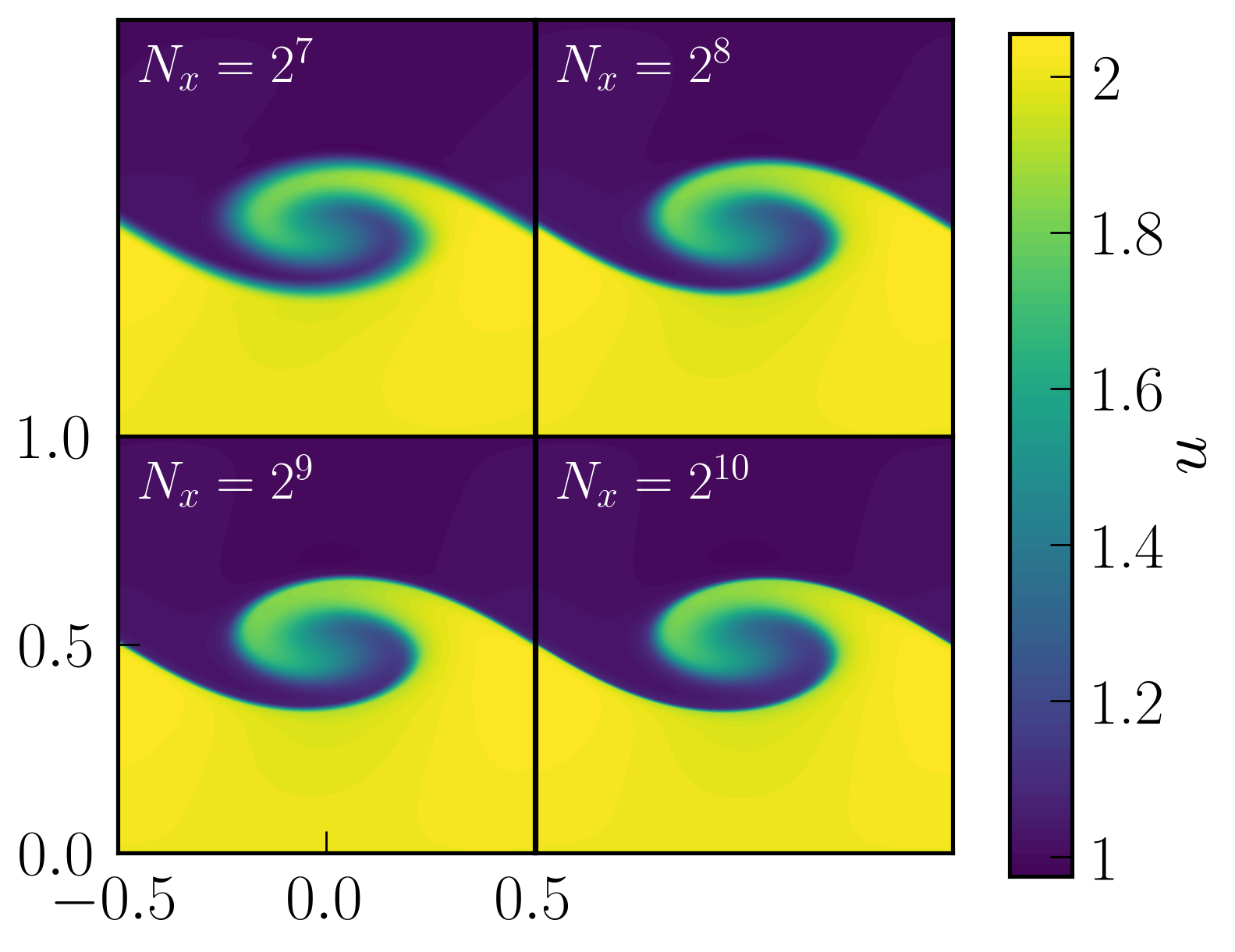}
    \caption{Snapshot of the density $n$ at $t = 11$ from the Kelvin-Helmholtz simulation for
    $\eta_0 = 4 \times 10^{-4}$, as a function of resolution.  For the two lower
    resolution panels (top row), the physical viscosity is small enough that
    the BDNK primitive variable solution is numerically unstable, and we use
    the adaptive primitive variable algorithm with tolerances $\Delta_\eta =
    10^{-3}, 10^{-4}$, respectively.  At these tolerances, the perfect fluid
    primitive variable solution is used across the entire grid for most of the
    simulation after the first few timesteps; despite this, the solution still
    converges to the correct viscous solution, and is noticeably different from
    the inviscid solution (top-left panel of Fig. \ref{fig:KH}) because most of
    the dissipation comes from the viscous fluxes rather than the primitive
    variable solution. 
  } \label{fig:kh_conv}
\end{figure}

Fig. \ref{fig:kh_conv} shows a set of screenshots at $t = 11$ of the $\eta_0 =
4 \times 10^{-4}$ simulations as a function of $N_{x}$.  At the lower two resolutions, the
BDNK primitive variable solver fails, and the solution can be stabilized using
the adaptive algorithm with $\Delta_{\eta} = 10^{-3}, 10^{-4}$ respectively.
In these cases, the perfect fluid primitive variable solution (\ref{eq:PF_pvr})
is used over essentially the entire grid.  Despite this fact, the solutions
produced by increasing resolution and shrinking the viscous tolerance
($\Delta_\eta$) still converge to the true BDNK viscous solution.  We stress
that most of the visible effect of viscosity is provided by the viscous fluxes,
which are numerically well-behaved in the inviscid limit; as a result, the top
two (lower resolution) panels of Fig. \ref{fig:kh_conv}, despite using the
perfect fluid primitive variable solution, still resemble (and converge to) 
the $\eta_0 = 4 \times 10^{-4}$ panel of Fig. \ref{fig:KH} rather than the $\eta_0 = 0$ 
panel.

\section{Conclusion} \label{sec:conclusion}
In this work we have presented the first multi-dimensional
finite volume scheme designed to
solve the causal, stable relativistic Navier-Stokes equations of BDNK
theory \cite{Bemfica:2020zjp}.  
Our algorithm naturally extends traditional HRSC central schemes
\cite{kurganov2000new,Lucas-Serrano:2004fqa}, but features distinct
modifications to account for the different structure of the BDNK
equations. In BDNK, the variables which must be
reconstructed also include spatial derivatives of the hydrodynamic
variables; to compute these in a non-oscillatory fashion, we use adaptive
derivative stencils based in the central-WENO (CWENO) method
\cite{Levy1999}. The most
involved difference between a finite volume perfect fluid solver and our
BDNK scheme comes in solving for the primitive variables (see, e.g.,
\cite{Noble:2005gf} for a discussion).
It turns out that the BDNK primitive variable solution may generically be carried out
analytically, though problems arise when viscous contributions are small
enough to become unresolved at a given grid resolution.  In these cases, we
apply an adaptive algorithm which treats cells with unresolved viscosity as
``effectively inviscid'', applying the perfect fluid primitive variable
inversion.  As resolution is increased, the viscous terms eventually become
resolved in these cells, and the BDNK primitive variable solution is used.
As a result, the adaptive approach produces solutions which converge to
solutions of the continuum PDEs.

To evaluate the algorithm's performance, we focus on the conformal
fluid limit and apply it to several
flat-spacetime test problems with variation in one and two spatial
dimensions.  We begin with a
simple test of smooth initial data in 1D, and confirm that the new scheme
conserves the integrals of motion up to machine precision. 
A 1D shock tube test
illustrates 
the improved stability of the new scheme over that of \cite{Pandya_2021}, and a 2D oblique shockwave test suggests a
higher-order time integrator may be useful to stably evolve very
high-velocity flows with sharp gradients.  Steady-state shockwave solutions
in 1D are used to illustrate the spatial dependence of the adaptive
primitive variable scheme, and 2D simulations demonstrating the
Kelvin-Helmholtz instability provide a case where the adaptive algorithm is
necessary to produce convergent low-viscosity BDNK solutions at finite
resolution.

While we have specialized to conformal fluids in this work, future
extensions will be equipped with more general microphysics, allowing for
the study of effects such as bulk viscosity. A generalization of the
approach presented here would also make possible a detailed comparison of
BDNK and MIS-type theories constructed in general hydrodynamic frames
\cite{Noronha:2021syv}, which could help elucidate the connections between
the two theories.  There are also a number of open numerical problems to be
investigated---one example would be to perform a comparison of fully
flux-conservative solvers for BDNK and MIS \cite{Most:2021rhr}; another would
be to consider a full first-order reduction of the BDNK equations, wherein
the spatial derivative terms are evolved using their own set of evolution
equations.

The BDNK algorithm presented here should be sufficiently stable and
accurate to be applied to a variety of relativistic hydrodynamics problems where
first-order dissipation might be relevant. 
Among those would be the investigation of viscous effects in the
inspiral \cite{Arras:2018fxj} and merger \cite{Most2021bulk} of binary
neutron star systems. The guaranteed causality of the BDNK equations
would also motivate the application of this numerical scheme to
simulations of heavy-ion collisions, where current MIS approaches show
acausal behavior \cite{Plumberg:2021bme}.
While viscous effects
might also be important in black-hole accretion problems
\cite{Chandra2017}, the presence of magnetic fields introduces
anisotropies in the dissipative sector presently unaccounted for in
BDNK theory \cite{Chandra:2015iza,Most:2021rhr}. Although first-order
formulations of dissipative MHD have been proposed \cite{Dommes:2020ktk}, their
extension to general hydrodynamic frames has just started to be
investigated \cite{Armas:2022wvb}.

\begin{acknowledgments}
The authors thank Fabio Bemfica, Marcelo Disconzi and Jorge Noronha for
insightful discussions related to this work.
This material is based upon work supported by the National Science
Foundation (NSF) Graduate Research Fellowship Program under Grant No.
DGE-1656466. Any opinions, findings, and conclusions or recommendations
expressed in this material are those of the authors and do not necessarily
reflect the views of the National Science Foundation. F.P. acknowledges
support from NSF Grant No.  PHY-1912171, the Simons Foundation, and the
Canadian Institute For Advanced Research (CIFAR). ERM acknowledges support
from postdoctoral fellowships at the Princeton Center for Theoretical
Science, the Princeton Gravity Initiative, and the Institute for Advanced
Study.
\end{acknowledgments}

\appendix

\section{Coordinate components of the conserved currents} \label{sec:coord_eqns}
Here we present the Cartesian components of the conserved currents $J^{a}_{1},
T^{ab}_{1}$ for a conformal fluid in 4D Minkowski spacetime.  Since we only
consider systems with translation invariance in the $z$ direction, only
the $t, x, y$ components will be necessary.  For 1D test problems (namely the
1D Gaussian, shock tube, and steady-state shockwave in Sec.
\ref{sec:numerical_tests}) we take all fields to only be functions of $t, x$,
and as a result $T^{cy}_{1} = u^{y} = 0$; these equations may also be found in
\cite{Pandya_2021}.

Beginning with the particle current (which is the same for the perfect
fluid and BDNK theory), the components of $J^{a}_{1}$ are obtained
immediately using (\ref{eq:J_cov}) and the four-velocity
\begin{equation}
u^{a} = \Big( \sqrt{1 + (u^{x})^2 + (u^{y})^2}, u^{x}, u^{y}, 0 \Big)^{T}.
\end{equation}
Combining the four velocity with the perfect fluid stress-energy tensor 
(\ref{eq:Tab_0_cov}) may be used compute $T^{ab}_{0}$ as well.

The BDNK stress-energy tensor may be specified by defining the components in
(\ref{eq:Tab_cov}).  The correction to the energy density is
\begin{multline}
\mathcal{A} = \frac{1}{4} \chi_0 e^{\frac{3}{4}\xi} \Big[u^{x} \left(\frac{4 \dot{u}^{x}}{u^{t}} + 3 \xi_{,x}\right) + u^{y} \left(\frac{4 \dot{u}^{y}}{u^{t}} + 3 \xi_{,y}\right) \\
+ 3 \dot{\xi} u^{t} + 4 u^{x}_{,x} + 4 u^{y}_{,y} \Big].
\end{multline}
The two independent components of the heat flux vector are
\begin{multline}
\mathcal{Q}^{x} = \frac{1}{4} \lambda_0 e^{\frac{3}{4} \xi} \Big[4 \dot{u}^{x} u^{t} + u^{x} \dot{\xi} u^{t} + \left((u^{x})^2+1\right) \xi_{,x} \\
+ 4 u^{x} u^{x}_{,x} + u^{y} (u^{x} \xi_{,y} + 4 u^{x}_{,y})\Big]
\end{multline}
and
\begin{multline}
\mathcal{Q}^{y} = \frac{1}{4} \lambda_0 e^{\frac{3}{4} \xi} \Big[ u^{y} (u^{t} \dot{\xi} + u^{x} \xi_{,x} + u^{y} \xi_{,y} + 4 u^{y}_{,y}) \\
+ 4 u^{t} \dot{u}^{y} + 4 u^{x} u^{y}_{,x} + \xi_{,y} \Big],
\end{multline}
which are related to the $t$ component by the constraint $u_{a} \mathcal{Q}^{a}
= 0$, which implies
\begin{equation}
\mathcal{Q}^{t} = \frac{1}{u^{t}} (u^{x} \mathcal{Q}^{x} + u^{y} \mathcal{Q}^{y}),
\end{equation}
and $\mathcal{Q}^{z} = 0$ due to the spatial symmetry we have assumed.  The independent
components of the shear term are the $xx$ component
\begin{multline}
-2 \eta \sigma^{xx} = \frac{2 \eta_0 e^{\frac{3}{4} \xi}}{3 u^{t}} \Big[ u^{t} \Big(-2 \left((u^{x})^2+1\right) u^{x}_{,x} + (u^{x})^2 u^{y}_{,y} \\
- 3 u^{x} u^{x}_{,y} u^{y} + u^{y}_{,y}\Big) - 2 (u^{x})^3 \dot{u}^{x} + (u^{x})^2 u^{y} \dot{u}^{y} \\
- u^{x} \dot{u}^{x} \left(3 (u^{y})^2 + 2\right) + u^{y} \dot{u}^{y} \Big],
\end{multline}
the $xy$ component
\begin{multline}
-2 \eta \sigma^{xy} = -\frac{\eta_0 e^{\frac{3}{4} \xi}}{3 u^{t}} \Big[ u^{t} \Big(3 (u^{x})^2 u^{y}_{,x}+u^{x} u^{y} (u^{x}_{,x}+u^{y}_{,y}) \\
+3 u^{x}_{,y} \left((u^{y})^2+1\right)+3 u^{y}_{,x}\Big) + \dot{u}^{x} u^{y} \left((u^{x})^2+3 (u^{y})^2+3\right)\\
+u^{x} \dot{u}^{y} \left(3 (u^{x})^2+(u^{y})^2+3\right) \Big],
\end{multline}
and the $yy$ component
\begin{multline}
-2 \eta \sigma^{yy} = \frac{2 \eta_0 e^{\frac{3}{4} \xi}}{3 u^{t}} \Big[-3 u^{t} u^{x} u^{y} u^{y}_{,x} + u^{t} u^{x}_{,x} \left((u^{y})^2+1\right) \\
- 2 u^{t} \left((u^{y})^2+1\right) u^{y}_{,y} - 3 (u^{x})^2 u^{y} \dot{u}^{y} + u^{x} \dot{u}^{x} \left((u^{y})^2+1\right) \\
- 2 \left((u^{y})^3+u^{y}\right) \dot{u}^{y}\Big].
\end{multline}
The remaining required components may be found from the constraint $u_{a}
\sigma^{ab} = 0$, which implies
\begin{equation}
\sigma^{t c} = \frac{1}{u^{t}} (u^{x} \sigma^{x c} + u^{y} \sigma^{y c}).
\end{equation}

\section{Primitive variable recovery for a non-conformal BDNK fluid} \label{sec:non_conformal_pvr}
For a non-conformal BDNK fluid, (\ref{eq:numerical_BDNK_q_1}) generalizes to
\begin{equation} \label{eq:q1_nonconformal}
\bm{q}_1 = \bm{q}_0(\bm{p}_0) + \bm{A}(\bm{p}_0) \cdot \bm{C} \cdot \bm{p}_1 + \bm{D} \cdot \bm{b}(\bm{p}_0, \partial_i \bm{p}_0) + \bm{\tau},
\end{equation}
where the matrices $\bm{C}, \bm{D}$ are populated solely with transport coefficients, and vanish in the inviscid limit.  In the conformal limit, $\bm{C}, \bm{D} \to \eta_0 \, \bm{I}$, where $\bm{I}$ is the identity matrix, recovering (\ref{eq:numerical_BDNK_q_1}).  The primitive variables may still be obtained analytically,
\begin{equation} \label{eq:p1_nonconformal}
\bm{p}_1 = \bm{C}^{-1} \cdot \bm{A}^{-1} \cdot \Big[ (\bm{q}_1 - \bm{q}_0) - \bm{D} \cdot \bm{b} - \bm{\tau} \Big],
\end{equation}
though (\ref{eq:p1_nonconformal}) suffers the same problems as its conformal analog in the inviscid limit, and all terms vanish except for $\bm{C}^{-1} \cdot A^{-1} \cdot \bm{\tau}$, which diverges at finite grid resolution.

To stabilize the scheme in these cases, one may compute $\bm{p}_1^{PF}$ using (\ref{eq:RE_eqns_nonconservative}) and compute a set of shifted variables $\tilde{\bm{q}}$, where (\ref{eq:q_tilde_defn}) generalizes to
\begin{equation} \label{eq:q_tilde_nonconformal}
\tilde{\bm{q}}_1 \equiv \bm{q}_1 - \bm{q}_1 \Big|_{\bm{p}_1 \to \bm{p}_1^{PF}} = \bm{A} \cdot \bm{C} \cdot (\bm{p}_1 - \bm{p}_1^{PF}),
\end{equation}
implying the solution $\bm{p}_1(\tilde{\bm{q}}_1)$ is
\begin{equation} \label{eq:p1_of_qtilde_nonconformal}
\bm{p}_1 = \bm{C}^{-1} \cdot \bm{A}^{-1} \cdot \tilde{\bm{q}}_1 + \bm{p}_1^{PF}.
\end{equation}
Assuming one suitably modifies (\ref{eq:trivial_evol_eqns}) to accommodate the choice of BDNK primitive variables, and one has a perfect fluid primitive variable solution for the case of interest\footnote{Note that in general the perfect fluid primitive variable solution is unobtainable analytically, so it is likely that a numerical solver will be necessary in cells where the physical viscosity is unresolved.  In cells where it is resolved, however, the analytic BDNK primitive solution (\ref{eq:p1_of_qtilde_nonconformal}) should be used.} (to replace (\ref{eq:PF_pvr})), one may use (\ref{eq:q_tilde_nonconformal}-\ref{eq:p1_of_qtilde_nonconformal}) in place of (\ref{eq:q_tilde_defn}-\ref{eq:shifted_BDNK_pvr}) in the algorithm described in Sec. \ref{sec:pvr} to obtain stable, convergent solutions to the BDNK equations in the inviscid limit.

\section{Review of WENO reconstruction} \label{sec:WENO}
For the sake of simplicity, we will review the WENO reconstruction algorithm
for a problem with variation only in one dimension; hence we will consider how
the algorithm constructs the primitive variables $p^{\pm}_{i+1/2}$ at the right ($+$) and left ($-$)
sides of the cell interface located at $x_{i+1/2} = x_{i} + \frac{1}{2} h$,
where $h$ is the grid spacing. Beginning with the reconstructed value at the
right side of the interface, $p^{+}_{i+1/2}$, the WENO algorithm begins with the
computation of three so-called ENO polynomials constructed from the cell
averages,
\begin{equation} \label{eq:WENO_right_stencils}
\begin{aligned}
v^{0}_{i+1/2} &= -\frac{1}{6} \bar{p}_{i-2} + \frac{5}{6} \bar{p}_{i-1} + \frac{1}{3} \bar{p}_{i} \\
v^{1}_{i+1/2} &= \frac{1}{3} \bar{p}_{i-1} + \frac{5}{6} \bar{p}_{i} - \frac{1}{6} \bar{p}_{i+1} \\
v^{2}_{i+1/2} &= \frac{11}{6} \bar{p}_{i} - \frac{7}{6} \bar{p}_{i+1} + \frac{1}{3} \bar{p}_{i+2}. \\
\end{aligned}
\end{equation}
Each of these stencils on its own constitutes an approximation to
$p^{+}_{i+1/2}$ that is third-order accurate in the grid spacing $h$.  WENO
achieves the essentially non-oscillatory property by adaptively weighting how
much of each stencil goes into the final approximation for $p^{+}_{i+1/2}$
using a set of \textit{smoothness indicators}
\begin{equation}
\begin{aligned}
\beta^0 &= \frac{1}{4} (3 \bar{p}_i - 4 \bar{p}_{i+1} + \bar{p}_{i+2})^2 + \frac{13}{12} (\bar{p}_i - 2 \bar{p}_{i+1} + \bar{p}_{i+2})^2 \\
\beta^1 &= \frac{1}{4} (\bar{p}_{i+1}-\bar{p}_{i-1})^2 + \frac{13}{12} (\bar{p}_{i-1} - 2 \bar{p}_{i} + \bar{p}_{i+1})^2 \\
\beta^2 &= \frac{1}{4} (\bar{p}_{i-2} - 4 \bar{p}_{i-1} + 3 \bar{p}_{i})^2 + \frac{13}{12} (\bar{p}_{i-2} - 2 \bar{p}_{i-1} + \bar{p}_{i})^2,
\end{aligned}
\end{equation}
where $\beta^{k}$ is large when the stencil $v^{k}_{i+1/2}$ contains a sharp
gradient.  Such stencils should have small weights in the final reconstructed
primitive variable, then, which is achieved by writing the weights $w_k$ as
\begin{equation} \label{eq:WENO_weights}
w_k = \frac{\alpha_k}{\sum_l \alpha_l}, ~~ \alpha_k = \frac{d_k}{(\epsilon_W + \beta^{k})^2}, ~~ d_k = \Big( \frac{3}{10}, \frac{3}{5}, \frac{1}{10} \Big)
\end{equation}
where the constant linear weights $d_k$ are chosen such that the reconstructed
solution attains the highest possible order of accuracy (5th order) when the
solution is smooth ($\beta^{k}$ is small) in all three stencils.  

The quantity $\epsilon_W$ is a free parameter which is inserted to prevent
divide-by-zero errors when the smoothness indicators $\beta^{k}$ vanish.
The sensitivity of the WENO algorithm to sharp features in the solution
depends strongly on the magnitude of $\epsilon_{W}$. Cases where $\epsilon_{W}$
is small can yield $w_k$ far from $d_k$ in nonsmooth regions, resulting in
significant differences between the stencils being used across the grid.
Conversely, the limit $\epsilon_{W} \to \infty$ forces $w_k \to d_k$,
recovering a fixed fifth-order reconstruction for $p^{\pm}_{i+1/2}$.

The final WENO approximation for $p^{+}_{i+1/2}$ is given by
\begin{equation} \label{eq:WENO_right_reconst}
p^{+}_{i+1/2} = w_0 v^0_{i+1/2} + w_1 v^1_{i+1/2} + w_2 v^2_{i+1/2},
\end{equation}
which, again, gives the value of $p$ at the right side of the interface at
$x_{i+1/2}$.  At the left side of the interface, the approximation is achieved
by reflecting the stencils (\ref{eq:WENO_right_stencils}) across the interface,
which yields ENO polynomials
\begin{equation} \label{eq:WENO_left_stencils}
\begin{aligned}
u^{0}_{i+1/2} &= \frac{1}{3} \bar{p}_{i} + \frac{5}{6} \bar{p}_{i+1} - \frac{1}{6} \bar{p}_{i+2} \\
u^{1}_{i+1/2} &= -\frac{1}{6} \bar{p}_{i-1} + \frac{5}{6} \bar{p}_{i} + \frac{1}{3} \bar{p}_{i+1} \\
u^{2}_{i+1/2} &= \frac{1}{3} \bar{p}_{i-2} - \frac{7}{6}\bar{p}_{i-1} + \frac{11}{6} \bar{p}_{i}. \\
\end{aligned}
\end{equation}
The smoothness indicators and linear weights (\ref{eq:WENO_weights}) remain the
same, giving the final approximation
\begin{equation} \label{eq:WENO_left_reconst}
p^{-}_{i+1/2} = w_0 u^0_{i+1/2} + w_1 u^1_{i+1/2} + w_2 u^2_{i+1/2}.
\end{equation}
For 2D simulations on uniform Cartesian grids like those considered here, WENO
reconstruction is applied in the same way in both spatial directions; to obtain
$p^{\pm}_{i,j+1/2}$, simply keep the index $i$ constant and swap $i \to j$ in
(\ref{eq:WENO_right_stencils}-\ref{eq:WENO_left_reconst}).

\section{Convergence tests}
\begin{figure*}
	\centering
	\includegraphics[width=0.85\textwidth]{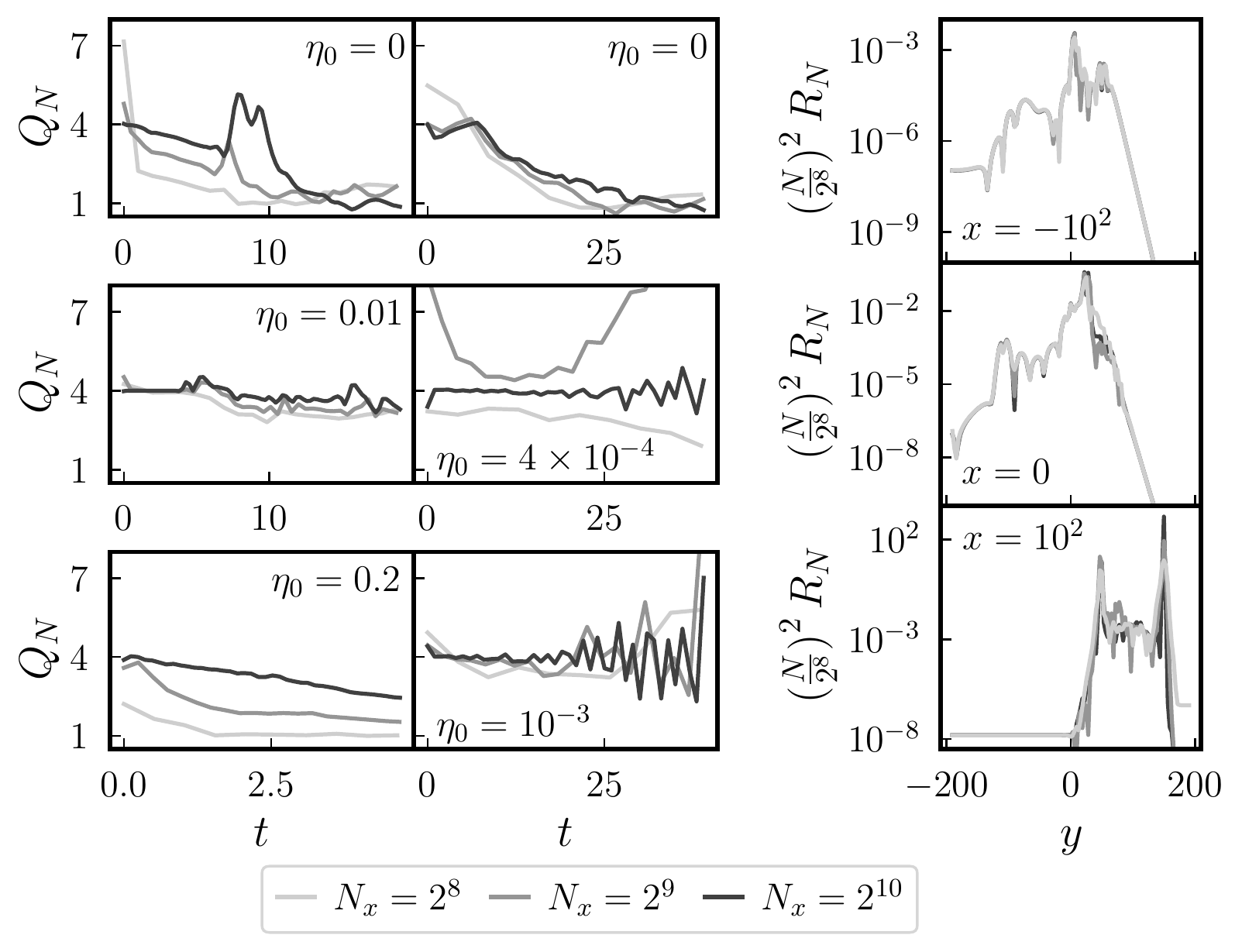}
    \caption{Convergence plots corresponding to an independent (Crank-Nicolson
    second-order finite difference) discretization of the $x$-component of
    (\ref{eq:T_cons_law}), $\nabla_c T^{cx} = 0$, for the 2D simulations shown
    above.  Leftmost column: plots of $Q_{N}(t)$ for the viscous rotor
    simulations showing that the curves approach second-order convergence as
    resolution increases (shown with finer grids in successively darker colors,
    corresponding to resolutions $N_x = 2^8, 2^9, 2^{10}$).  Middle column:
    similar plots of $Q_{N}(t)$ for the Kelvin-Helmholtz-unstable initial data.
    Rightmost panel: plots of the residual $\nabla_c T^{cx}$ for the 2D oblique
    shock initial data, at the time shown in Fig. \ref{fig:2D_step}, with the
    same color coding by resolution, scaled such that all curves should overlap
    if they are converging at second order, e.g. the $N_x = 2^{9}$ curve is
    multiplied by $4$ and the $N_x = 2^{10}$ curve is multiplied by 16.  The
    three panels show successive slices through the domain at constant $x$, and
    the top two show convergence at the expected order (all curves overlap).
    The bottom curve shows a slice through the shockwave, and converges roughly
    at the expected order everywhere outside the spikes which appear at the
    shock fronts.  Increasing resolution should produce taller, thinner spikes
    at the shockwave until it is finally resolved and the solution begins
    converging there at second order.} \label{fig:conv_plot}
\end{figure*}

To check the validity of the simulation results presented here, we have
performed a number of different convergence tests.  Principal among these is
the computation of a set of independent residuals $R_{N}$, which are copies of
the equations of motion (\ref{eq:T_cons_law}-\ref{eq:J_cons_law}) with
different discretizations from those solved in the solution algorithm.  For a
second-order accurate numerical scheme (as presented here), a
second-order-accurate independent residual should yield $R_{N} \propto O(h^2)$,
where the grid spacing $h$ is related to the number of gridpoints $N_{x}$ by $h
= \frac{L_+ - L_-}{N_{x}}$, and $L_\pm$ are the left and right sides of the
domain in either spatial coordinate (as we use equal grid spacing in $x$ and
$y$).  Hence, a quantitative measure of convergence would be to compute the
instantaneous convergence factor
\begin{equation}
Q_{N}(t) = \frac{|| R_{N/2} ||}{|| R_{N} ||},
\end{equation}
where $|| \cdot ||$ is any vector norm.  Here we use the $1$-norm, defined for
a vector $X$ to be $||X||_{1} \equiv \sum_{i} |X_i|$.
For a second-order-accurate scheme,
when the solution is sufficiently smooth, one can show that $Q_{N}(t) \to 4$ as
$h \to 0$.

The convergence factor $Q_{N}(t)$ implicitly assumes that the solution is
smooth everywhere, which is not the case here when unresolved shockwaves 
are present.
In these cases, we instead directly compute the independent residuals $R_{N}$
over the spatial grid, and confirm that these obey $R_N \propto O(h^2)$ in
smooth regions.  An unresolved shockwave appears as step function
discontinuity, which in the independent residual (which typically involves
derivatives across the discontinuity) generates a delta function-like peak
which grows taller and narrower with resolution.  We confirm that this behavior
occurs at the unresolved shockwaves present in our simulations.

Fig. \ref{fig:conv_plot} shows the convergence of an independent
residual---namely, a second-order Crank-Nicolson finite difference
discretization of the $x$-component of (\ref{eq:T_cons_law})---for the viscous
rotor, Kelvin-Helmholtz, and 2D oblique shockwave simulations in columns from
left to right.  For the left two columns, the solutions are free of unresolved shockwaves
and $Q_{N}(t)$ is a good measure of convergence; for each of the cases shown
(which differ in viscosity), the solutions approach $Q_N(t) = 4$ with
increasing resolution (which is denoted with lines of increasing darkness).
Note that the inviscid simulations lose convergence at late times; this is
because they begin forming features at the grid scale, which are unresolved at
lower resolution.  The time at which convergence is lost, however, is pushed
later and later as resolution increases, as expected.  The third column of the
figure shows the independent residual $R_{N}$ on constant-$x$ slices at $t = 220$
as in Fig. \ref{fig:2D_step}, as a
function of $y$, scaled by the expected order of convergence such that the
three lines should overlap if the scheme is converging at second order.  One
can see that the top two slices exhibit the expected order of convergence, and
all three resolutions lie on top of each other; in the bottom plot (which
passes through the shockwaves), delta-function-like spikes form at the two
shock fronts, and grow taller and narrower with resolution, as expected.
Elsewhere the solutions roughly converge at the expected rate, though the sharp
gradients in this region produce some numerical ``noise'' as well.

Similar trends to those shown in Fig. \ref{fig:conv_plot} appear in independent
residuals of the other components of (\ref{eq:T_cons_law}); computations of the
self-convergence of the hydrodynamic variables $\{\xi, n, u^x, u^y\}$ are even
better-behaved, and converge at second order as well.

\bibliography{references}% Produces the bibliography via BibTeX.

\end{document}